\documentclass{vldb}
\usepackage{graphicx}
\usepackage{balance}  
\usepackage{listings}
\usepackage[table]{xcolor}
\usepackage[linesnumbered, ruled,vlined]{algorithm2e}
\usepackage{hyperref} 
\usepackage{flushend}
 
\definecolor{dkgreen}{rgb}{0,0.6,0}
\definecolor{gray}{rgb}{0.5,0.5,0.5}
\definecolor{mauve}{rgb}{0.58,0,0.82}
\definecolor{blue}{rgb}{0.15, 0.36, 0.7}
 
\lstdefinestyle{mystyle}{
  language=Python,
  aboveskip=3mm,
  belowskip=3mm,
  showstringspaces=false,
  basicstyle={\footnotesize\ttfamily},
  xleftmargin=2em,
  numbers=left,
  basicstyle=\ttfamily\scriptsize,
  numberstyle=\tiny\color{gray},
  keywordstyle=\color{blue},
  commentstyle=\color{dkgreen},
  stringstyle=\color{red},
  breaklines=true,
  breakatwhitespace=true,
  tabsize=4,
}
 
\def\code#1{\texttt{{\footnotesize#1}}}

\begin{document}


\title{PyTorch Distributed: Experiences on Accelerating \\ Data Parallel Training}

\author{Shen Li$^{\textbf{\dag}}$ \ \ Yanli Zhao$^{\textbf{\dag}}$ \ \ Rohan Varma$^{\textbf{\dag}}$ \ \ Omkar Salpekar$^{\textbf{\dag}}$ \\ Pieter Noordhuis\thanks{This work was conducted when Pieter Noordhuis was an employee at Facebook.} \ \ Teng Li$^{\textbf{\dag}}$ \ \ Adam Paszke$^{\textbf{\ddag}}$ \\ Jeff Smith$^{\textbf{\dag}}$ \ \ Brian Vaughan$^{\textbf{\dag}}$ \ \ Pritam Damania$^{\textbf{\dag}}$ \ \ Soumith Chintala$^{\textbf{\dag}}$ \\ \\
\{shenli, yanlizhao, rvarm1, osalpekar\}@fb.com, \\ pcnoordhuis@gmail.com, \ \ tengli@fb.com, \ \  adam.paszke@gmail.com, \\ \{jeffksmith, bvaughan, pritam.damania, soumith\}@fb.com \\ \\
$^{\textbf{\dag}}$Facebook AI \qquad \qquad $^{\textbf{\ddag}}$University of Warsaw}

\maketitle

\begin{abstract}

This paper presents the design, implementation, and evaluation of the PyTorch distributed data parallel module. PyTorch is a widely-adopted scientific computing package used in deep learning research and applications. Recent advances in deep learning argue for the value of large datasets and large models, which necessitates the ability to scale out model training to more computational resources. Data parallelism has emerged as a popular solution for distributed training thanks to its straightforward principle and broad applicability. In general, the technique of distributed data parallelism replicates the model on every computational resource to generate gradients independently and then communicates those gradients at each iteration to keep model replicas consistent. Despite the conceptual simplicity of the technique, the subtle dependencies between computation and communication make it non-trivial to optimize the distributed training efficiency. As of v1.5, PyTorch natively provides several techniques to accelerate distributed data parallel, including bucketing gradients, overlapping computation with communication, and skipping gradient synchronization. Evaluations show that, when configured appropriately, the PyTorch distributed data parallel module attains near-linear scalability using 256 GPUs.

\end{abstract}
\newpage
\section{Introduction}

Deep Neural Networks (DNN) have powered a wide spectrum of applications, ranging from image recognition~\cite{resnet}, language translation~\cite{bert}, anomaly detection~\cite{du2017deeplog}, content recommendation~\cite{van2013deep}, to drug discovery~\cite{ramsundar2019deep}, art generation~\cite{mao2017deepart}, game play~\cite{guo2014deep}, and self-driving cars~\cite{bojarski2016end}. Many applications pursue higher intelligence by optimizing larger models using larger datasets, craving advances in distributed training systems. Among existing solutions, distributed data parallel is a dominant strategy due to its minimally intrusive nature. This paper presents the design, implementation, and evaluation of the distributed data parallel package in PyTorch v1.5~\cite{pytorch:nips:2019}.

Training a DNN model usually repeatedly conducts three steps~\cite{lecun1988theoretical}, the forward pass to compute loss, the backward pass to compute gradients, and the optimizer step to update parameters. The concept of data parallelism is universally applicable to such frameworks. Applications can create multiple replicas of a model, with each model replica working on a portion of training data and performing the forward and backward passes independently. After that, model replicas can synchronize either their gradients or updated parameters depending on the algorithm. It's nominally possible to build a working version of data parallel purely on the application side, as it only requires inserting appropriate communications into every iteration. However, squeezing out the last bit of performance takes an enormous amount of effort in design and tuning. Providing native distributed data parallel APIs on the platform side would help application developers focus on optimizing their models, while the platform developing team could continuously and transparently improve the training speed. To provide a general distributed data parallel package, the challenges are three-fold.
\begin{itemize}
    \item \textbf{Mathematical equivalence}: The purpose of data parallel is to speed up training on large datasets. Applications expect to harvest the same result model as if all training had been performed locally without model replication. This requires mathematical equivalence to local training despite its distributed nature.
    \item \textbf{Non-intrusive and interceptive API}: Application developments usually start from local models and then scale out when necessary. To avoid the exorbitant hurdles during the transition, the API must be non-intrusive in application code. On the other hand, the API needs to allow the internal implementation to timely \emph{intercept} signals to carry out communications and system optimizations. 
    \item \textbf{High Performance}: Data parallel training is subject to subtle dependencies between computations and communications. The design and implementation have to explore the solution space to efficiently convert more resources into higher training throughput.
\end{itemize}
PyTorch provides distributed data parallel as an \code{nn.Module} class, where applications provide their model at construction time as a sub-module. To guarantee mathematical equivalence, all replicas start from the same initial values for model parameters and synchronize gradients to keep parameters consistent across training iterations. To minimize the intrusiveness, the implementation exposes the same \code{forward}~\cite{forward} API as the user model, allowing applications to seamlessly replace subsequent occurrences of a user model with the distributed data parallel model object with no additional code changes. Several techniques are integrated into the design to deliver high-performance training, including bucketing gradients, overlapping communication with computation, and skipping synchronization.

Evaluations were conducted on an exclusive 32-GPU cluster and on 256 GPUs from a much larger shared entitlement. We developed benchmarks to evaluate the distributed package across different scales to present an in-depth view of the performance implications of different optimization techniques and configurations. Experiments also cover the comparison between NCCL and Gloo communication libraries. The results show that 1) communication is the dominant training latency contributor, and its impact increases with model sizes; 2) bucket sizes considerably affect communication efficiency, which could lead to more than 2X speedup if configured properly; 3) skipping synchronizations appropriately would significantly reduce amortized communication overhead without noticeably degrading convergence speed.

Techniques described in this paper were first released in PyTorch v1.1. During the past year, we have seen significant adoption both internally and externally. %
Within Facebook, a workload study from \code{05/11/20} to \code{06/05/20} shows that more than 60\% of production GPU hours during that period were spent on the PyTorch distributed data parallel package across a wide variety of applications, including speech, vision, mobile vision, translation, etc. 
%
There are three main contributions in this paper. First, this paper reveals the design and implementation of a widely adopted industrial state-of-the-art distributed training solution. Second, this paper highlights real-world caveats (\emph{e.g.}, due to pluralized graphs) that were overlooked by prior work. Third, we share performance tuning experiences collected from serving internal teams and open-source community users and summarized several directions for future improvements.

The remainder of the paper is organized as follows. Section~\ref{sec:background} briefly introduces PyTorch and data parallelism. Section~\ref{sec:design} elaborates the design for the PyTorch distributed data parallel module. Implementations and evaluations are presented in Section~\ref{sec:impl} and Section~\ref{sec:eval} respectively. Then, Section~\ref{sec:discussion} discusses lessons learned and opportunities for future improvements, and Section~\ref{sec:related} surveys related work. Finally, Section~\ref{sec:conclusion} concludes the paper.

\newpage
\section{Background} \label{sec:background}

Before diving into distributed training, let us briefly discuss the implementation and execution of local model training using PyTorch. Then, we explain and justify the idea of data parallelism and describe communication primitives.

\subsection{PyTorch}

PyTorch organizes values into \code{Tensor}s which are generic n-dimensional arrays with a rich set of data manipulating operations. A \code{Module} defines a transform from input values to output values, and its behavior during the forward pass is specified by its \code{forward} member function. A \code{Module} can contain \code{Tensor}s as parameters. For example, a \code{Linear} \code{Module} contains a \code{weight} parameter and a \code{bias} parameter, whose \code{forward} function generates the output by multiplying the input with the \code{weight} and adding the \code{bias}. An application composes its own \code{Module} by stitching together native \code{Module}s (\emph{e.g.}, linear, convolution, etc.) and \code{Function}s (\emph{e.g.}, relu, pool, etc.) in the custom \code{forward} function. A typical training iteration contains a forward pass to generate losses using inputs and labels, a backward pass to compute gradients for parameters, and an optimizer step to update parameters using gradients. More specifically, during the forward pass, PyTorch builds an autograd graph to record actions performed. Then, in the backward pass, it uses the autograd graph to conduct backpropagation to generate gradients. Finally, the optimizer applies the gradients to update parameters. The training process repeats these three steps until the model converges.

\subsection{Data Parallelism}

PyTorch offers several tools to facilitate distributed training, including \code{DataParallel} for single-process multi-thread data parallel training using multiple GPUs on the same machine, \code{DistributedDataParallel} for multi-process data parallel training across GPUs and machines, and \code{RPC}~\cite{rpc} for general distributed model parallel training (\emph{e.g.}, parameter server~\cite{ps}). This paper focuses on \code{DistributedDataParallel}. Data parallelism enables distributed training by communicating gradients before the optimizer step to make sure that parameters of all model replicas are updated using exactly the same set of gradients, and hence model replicas can stay consistent across iterations. 

Parameter averaging is another popular technique to scale out model training. Similarly, it can launch multiple processes across multiple machines, but instead of synchronizing gradients, parameter averaging directly computes the average of all model parameters. This occurs after the local optimizer step, meaning that parameter averaging can be implemented completely as an auxiliary step and does not need to interact with local training steps at all, which is attractive as it can easily and cleanly decouple the code of distributed training and local iterations. There are several caveats with parameter averaging. 

\begin{itemize}
    \item Parameter averaging can produce vastly different results compared to local training, which, sometimes, can be detrimental to model accuracy. The root cause is that parameter averaging is \textbf{not} mathematically equivalent to processing all input data locally, especially when the optimizer relies on past local gradients values (\emph{e.g.}, momentum). As different model replicas are likely to see different gradients, the states in optimizers can gradually diverge, causing conflicting gradient descent directions. This can result in inexplicable differences in performance when switching from locally optimized models to large scale deployed models. 
    \item The structure of parameter averaging orchestrates computation (\emph{i.e.}, backward pass) and communication (\emph{i.e.}, computing average) into non-overlapping phases, using optimizer \code{step()} functions as a hard separation point. Regardless of how vigorously we optimize the computation or communication, one type of resource will stay idle at any given time instance, giving up a substantial performance optimization opportunity. 
\end{itemize}

Given the above fundamental pitfalls, we decided to implement distributed training using data parallelism to synchronize gradients instead of parameters. Note that, applications can still easily build parameter averaging using PyTorch. In fact, the collective communication feature described in Section~\ref{sec:collective} is an appropriate solution for this use case. Applications just need to explicitly launch \code{AllReduce} operations to calculate averaged parameters accordingly.

\subsection{AllReduce}

\code{AllReduce} is the primitive communication API used by \code{DistributedDataParallel} to compute gradient summation across all processes. It is supported by multiple communication libraries, including NCCL~\cite{nccl}, Gloo~\cite{gloo}, and MPI~\cite{mpi}. The \code{AllReduce} operation expects each participating process to provide an equally-sized tensor, collectively applies a given arithmetic operation (\emph{e.g.}, \code{sum}, \code{prod}, \code{min}, \code{max}) to input tensors from all processes, and returns the same result tensor to each participant. A naive implementation could simply let every process broadcast its input tensor to all peers and then apply the arithmetic operation independently. However, as \code{AllReduce} has significant impact on distributed training speed, communication libraries have implemented more sophisticated and more efficient algorithms, such as ring-based \code{AllReduce}~\cite{nccl} and tree-based \code{AllReduce}~\cite{nccl:tree_reduce}. As one \code{AllReduce} operation cannot start until all processes join, it is considered to be a synchronized communication, as opposed to the P2P communication used in parameter servers~\cite{ps}.
\section{System Design}\label{sec:design}

\begin{figure}
\centering
  \includegraphics[width=0.5\linewidth]{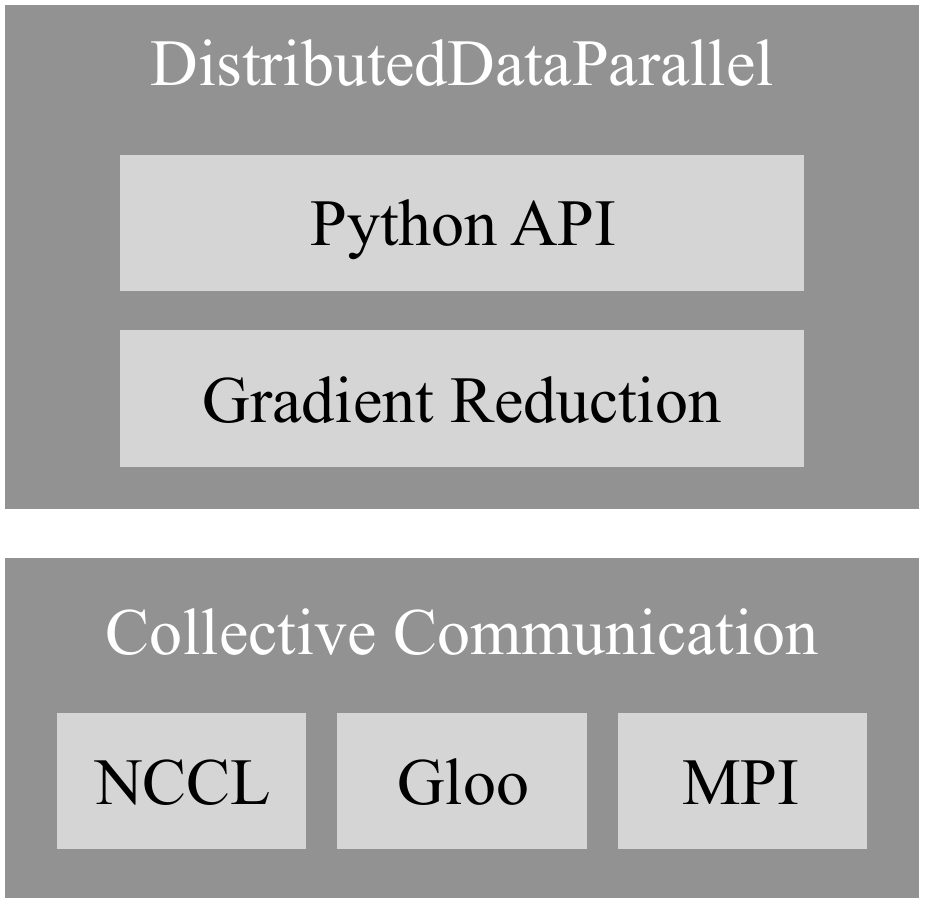}
  \caption{DistributedDataParallel Building Blocks}\label{fig:block}
  \vspace{-1.5em}
\end{figure}

PyTorch~\cite{pytorch:nips:2019} provides a \code{DistributedDataParallel} (\code{DDP}\footnote{For simplicity, the rest of the paper uses the acronym \code{DDP} to represent \code{DistributedDataParallel} henceforth.}) module to help easily parallelize training across multiple processes and machines. During distributed training, each process has its own local model replica and local optimizer. In terms of correctness, distributed data parallel training and local training must be mathematically equivalent. \code{DDP} guarantees the correctness by making sure that all model replicas start from the exact same model state, and see the same parameter gradients after every backward pass. Therefore, even though optimizers from different processes are all independent, they should be able to bring their local model replicas to the same state at the end of every iteration\footnote{For optimizers with intrinsic randomness, different processes can initialize their states using the same random seed.}. Fig.~\ref{fig:block} illustrates building blocks of \code{DDP}, which contains a Python API frontend, C++ gradient reduction core algorithm, and employs the \code{c10d} collective communication library. The following sections are presented in the top-down order of this stack graph. 

Section~\ref{sec:api} presents API design principles. Section~\ref{sec:reduction} explains gradient reduction techniques used in PyTorch distributed data parallel training. Finally, Section~\ref{sec:collective} discusses the collective communication backends for \code{DDP}.

\subsection{API}\label{sec:api}

When designing the API, we have defined two design goals to achieve the necessary functionality.

\begin{itemize}
    \item \textbf{Non-intrusive}: The API must be non-intrusive to applications. Application developers usually start from writing local training scripts, and scale out when hitting the resource limit on a single machine. At that point, it is unacceptable to ask developers to rewrite the entire application to enable distributed data parallel training. Instead, the developer should be able to reuse the local training script with minimal modifications.
    \item \textbf{Interceptive}: The API needs to allow the implementation to intercept various signals and trigger appropriate algorithms promptly. Distributed data parallel aims at accelerating training by using more computational resources. This process requires subtle optimizations in both computations and communications to achieve the best performance. Hence, the API must expose as many optimization opportunities as possible to the internal implementation.
\end{itemize}

Given the above requirements, we implemented distributed data parallel as an \code{nn.Module}, which takes the local model as a constructor argument and transparently synchronizes gradients in the backward pass. The code snippet below shows an example of using \code{DDP} module. This  example uses an \code{nn.Linear} layer to create a local model on line 10. Then, it converts the local model into a distributed training model on line 11 and sets up the optimizer on line 12. Line 14 through 23 are typical forward pass, backward pass, and optimizer step implementations. In this toy distributed training example, line 11 is the only difference that converts a local training application into a distributed one, which satisfies the non-intrusive requirement. It also fulfills the interceptive requirement. The constructor allows \code{DDP} to inspect the model structure and parameters. After construction, the local model is replaced by the distributed one, which can then easily intercept the \code{forward()} call to perform necessary actions accordingly. For the backward pass, \code{DDP} relies on backward hooks to trigger gradient reduction, which will be invoked by the autograd engine when executing \code{backward()} on the loss tensor. 


\begin{lstlisting}[language=Python]
import torch
import torch.nn as nn
import torch.nn.parallel as par
import torch.optim as optim

# initialize torch.distributed properly 
# with init_process_group

# setup model and optimizer
net = nn.Linear(10, 10)
net = par.`\textcolor{red}{DistributedDataParallel}`(net)
opt = optim.SGD(net.parameters(), lr=0.01)

# run forward pass
inp = torch.randn(20, 10)
exp = torch.randn(20, 10)
out = net(inp)

# run backward pass
nn.MSELoss()(out, exp).backward()

# update parameters
opt.step()
\end{lstlisting}
\vspace{-1em}

\begin{figure}
\begin{minipage}[c]{0.95\linewidth}
\begin{minipage}[c]{0.49\textwidth}
  \centering
  \includegraphics[width=\linewidth]{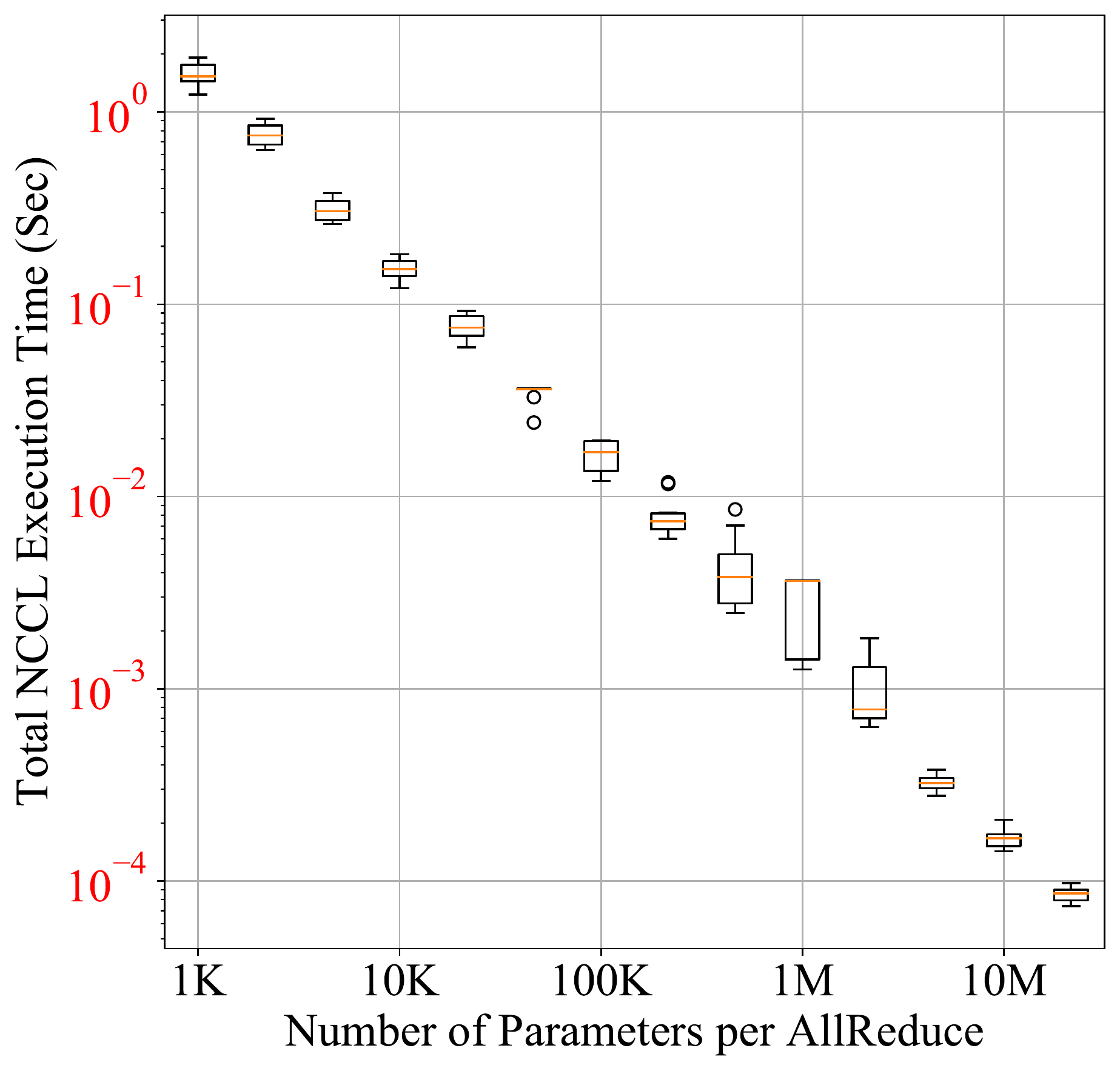}
  (a) NCCL
\end{minipage}
\begin{minipage}[c]{0.49\textwidth}
  \centering
  \includegraphics[width=\linewidth]{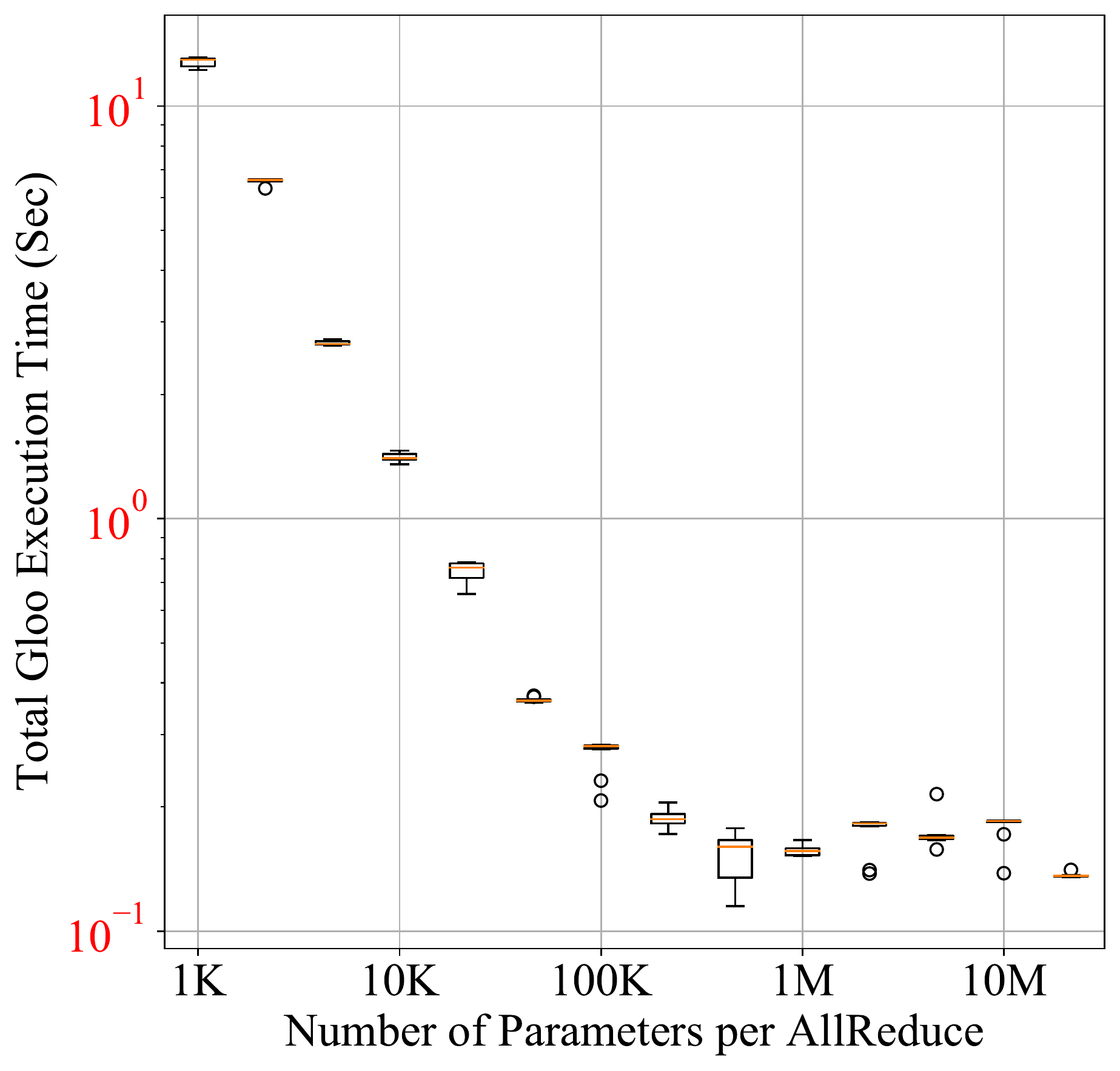}
  (b) GLOO
\end{minipage}
\end{minipage}
\begin{minipage}[c]{0.95\linewidth}
\begin{minipage}[c]{0.49\textwidth}
  \centering
  \includegraphics[width=\linewidth]{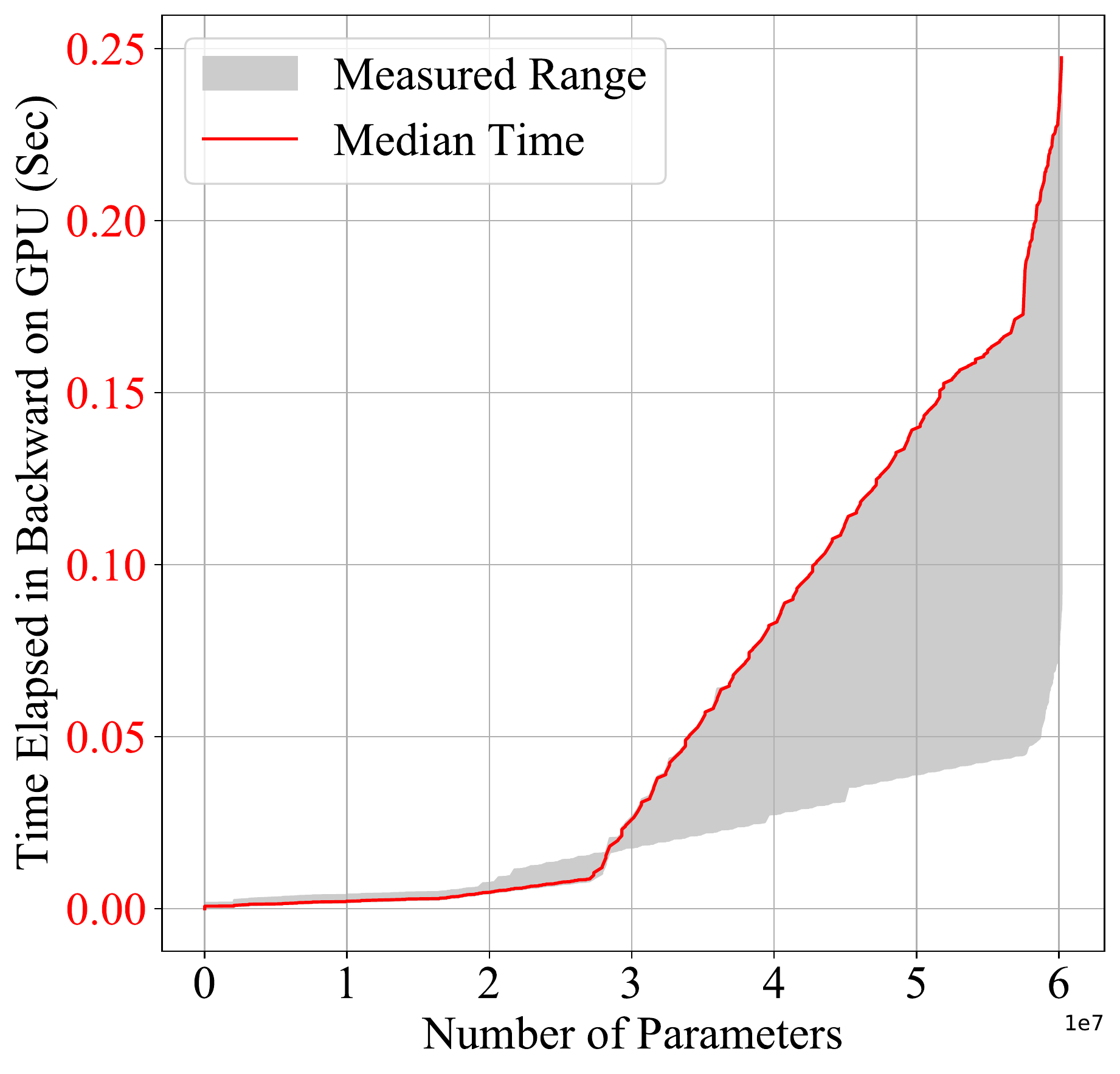}
  (c) GPU
\end{minipage}
\begin{minipage}[r]{0.49\textwidth}
\flushright
\begin{minipage}[c]{0.96\textwidth}
  \centering
  \includegraphics[width=\linewidth]{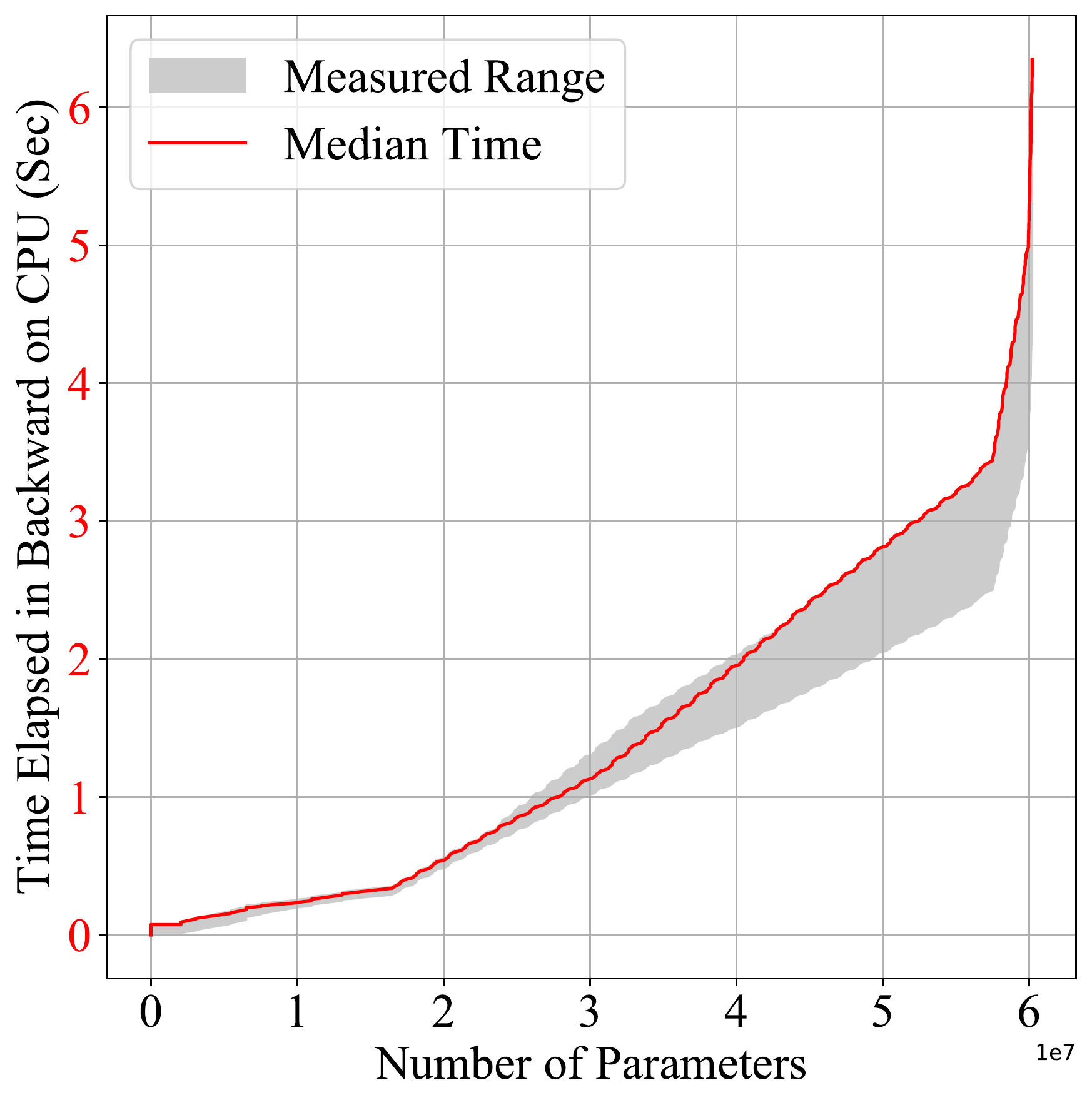}
  (d) CPU
\end{minipage}
\end{minipage}
\end{minipage}
\vspace{-1em}
\caption{Communication vs Computation Delay}\label{fig:allreduce}
\vspace{-1em}
\end{figure}

\subsection{Gradient Reduction}\label{sec:reduction}

The gradient reduction algorithm in \code{DDP} has evolved over the past releases. To introduce the structure of the current implementation, let us start from a naive solution, gradually introduce more complexities, and land in the current version as of today in PyTorch v1.5.0. This will also explain how the same simple API described in Section~\ref{sec:api} allows us to install various performance optimization algorithms.

\subsubsection{A Naive Solution}

As mentioned in the beginning of Section~\ref{sec:design}, \code{DDP} guarantees correctness by letting all training processes (1) start from the same model state and (2) consume the same gradients in every iteration. The former can be achieved by broadcasting model states from one process to all others at the construction time of \code{DDP}. To implement the latter, a naive solution can insert a gradient synchronization phase after the local backward pass and before updating local parameters. However, the API shown in Section~\ref{sec:api} does not provide an explicit entry point for this phase as there is nothing between \code{backward()} and \code{step()}. Fortunately, the PyTorch autograd engine accepts custom backward hooks. \code{DDP} can register autograd hooks to trigger computation after every backward pass. When fired, each hook scans through all local model parameters, and retrieves the gradient tensor from each parameter. Then, it uses the \code{AllReduce} collective communication call to calculate the average gradients on each parameter across all processes, and writes the result back to the gradient tensor.

The naive solution is sufficient for our purposes, but there are two performance concerns. 
\vspace{-0.3em}
\begin{itemize}
    \item Collective communication performs poorly on small tensors, which will be especially prominent on large models with massive numbers of small parameters. 
    \vspace{-0.3em}
    \item Separating gradient computation and synchronization forfeits the opportunity to overlap computation with communication due to the hard boundary in between.
\end{itemize}
The following sections elucidates solutions to address the above two concerns.

\subsubsection{Gradient Bucketing}

The idea of gradient bucketing is motivated by the observation that collective communications are more efficient on large tensors. Fig.~\ref{fig:allreduce} (a) and (b) provide a quantitative view, which show the total execution time to \code{AllReduce} 60M \code{torch.float32} parameters with different numbers of parameters per \code{AllReduce}. To maximize the bandwidth utilization, \code{AllReduce} operations are launched asynchronously and block waiting on all of them together, mimicking \code{DDP}'s gradient reduction algorithm. The experiments are conducted on an NVLink~\cite{nvlink} enabled server with two NVIDIA Quadro GP100 GPUs. NCCL~\cite{nccl} \code{AllReduce} runs on CUDA input tensors directly, while Gloo~\cite{gloo} \code{AllReduce} runs on CPU input tensors to eliminate the overhead of copying between CUDA memory to CPU memory when using Gloo backend. The total communication time clearly decreases when using larger input tensors, for both NCCL and Gloo. Gloo reaches pinnacle speed at around 500K parameters per input tensor, while there is no clear saturation signal for NCCL on NVLink with even 20M-parameter GPU tensors. 

These experiments suggest that, instead of launching a dedicated \code{AllReduce} immediately when each gradient tensor becomes available, \code{DDP} can achieve higher throughput and lower latency if it waits for a short period of time and buckets multiple gradients into one \code{AllReduce} operation. This would be especially helpful for models with many small parameters. However, \code{DDP} should not communicate all gradients in one single \code{AllReduce}, otherwise, no communication can start before the computation is over. Fig.~\ref{fig:allreduce} (c) and (d) show the GPU and CPU backward computations time of a ResNet152~\cite{resnet} that contains roughly 60M parameters. The X axis is the number of ready gradients and the Y axis the time elapsed since the beginning of the backward pass. The backward on GPU takes about 250ms to complete, which is in the same order of magnitude as NCCL on NVLink. This conclusion also applies to Gloo and CPU backward. These measurements herald that, with relatively small bucket sizes, \code{DDP} can launch \code{AllReduce} operations concurrently with the backward pass to overlap communication with computation, which would make a difference in per iteration latency. 

\subsubsection{Overlap Computation with Communication}\label{sec:overlap}

The \code{AllReduce} operation on gradients can start before the local backward pass finishes. With bucketing, \code{DDP} only needs to wait for all contents in the same bucket before launching communications. Under such settings, triggering \code{AllReduce} at the end of the backward pass is no longer sufficient. It needs to react to more frequent signals and launches \code{AllReduce} more promptly. Therefore, \code{DDP} registers one autograd hook for each gradient accumulator. The hook fires after its corresponding accumulator updating the gradients, and will inspect the bucket it pertains. If hooks of all gradients in the same buckets have fired, the last hook will trigger an asynchronous \code{AllReduce} on that bucket. 

\begin{figure}
\centering
  \includegraphics[width=0.9\linewidth]{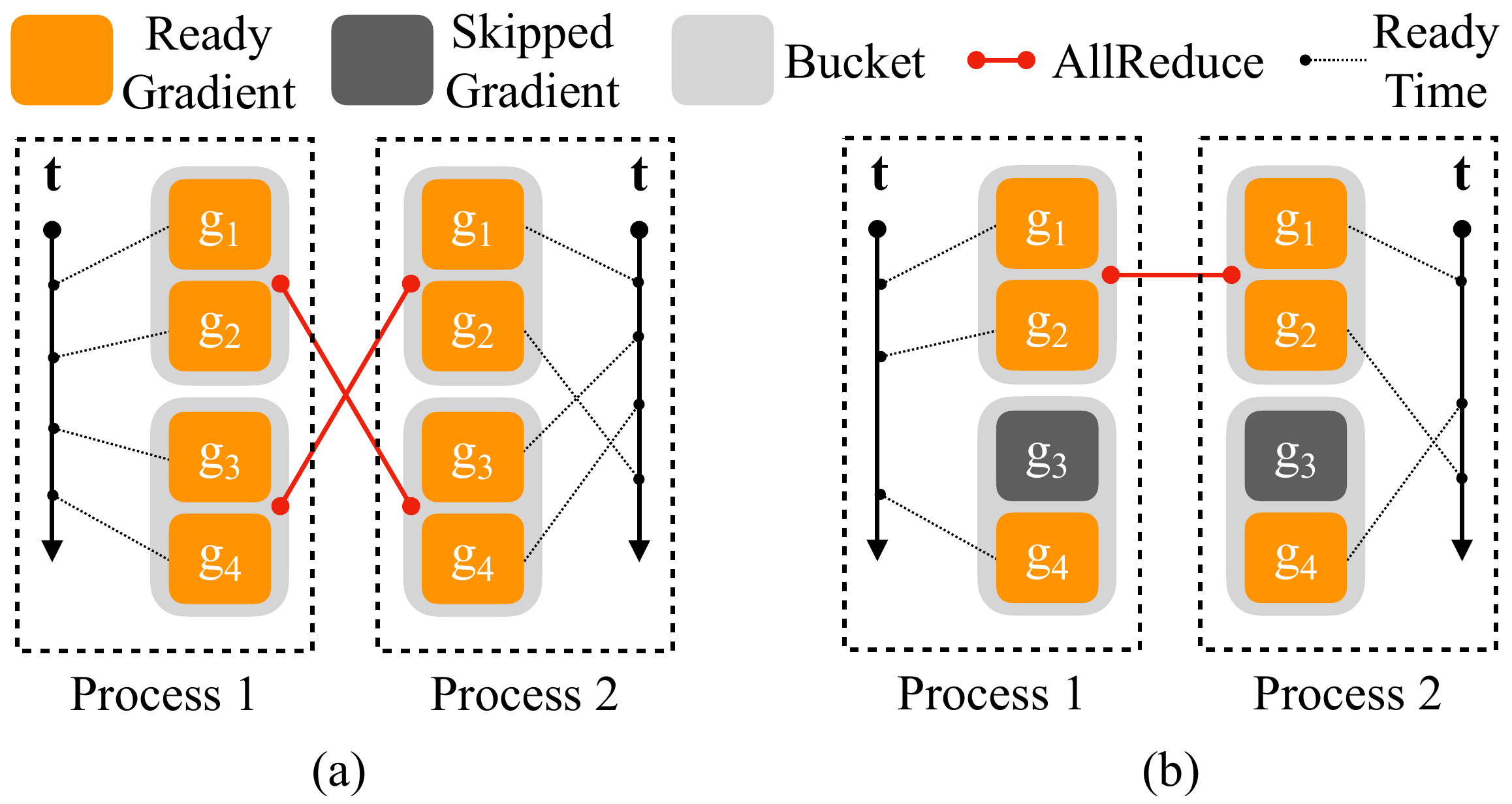}
  \vspace{-1em}
  \caption{Gradient Synchronization Failures}\label{fig:caveats}
  \vspace{-1.5em}
\end{figure}

Two caveats require caution. First, the reducing order must be the same across all processes, otherwise, \code{AllReduce} contents might mismatch, resulting in incorrect reduction result or program crash. However, PyTorch dynamically builds the autograd graph in every forward pass, and different processes might not agree on the gradient ready order. Fig.~\ref{fig:caveats}~(a) shows one example, where the two vertical axes represent time and dotted lines indicate when a gradient is ready. In process 1, the four gradients are computed in order, but the gradient $g_2$ are computed after $g_3$ and $g_4$ on process 2. In this case, if all processes \code{AllReduce} buckets as soon as they become ready, the \code{AllReduce} content would mismatch. Therefore,
all processes must use the same bucketing order, and no process can launch \code{AllReduce} on bucket \code{i+1} before embarking bucket \code{i}. If bucket \code{0} is the last one that becomes ready, there is no way that communication can overlap with computation. PyTorch v1.5.0 addresses this problem by using the reverse order of \code{model.parameters()} as the bucketing order, assuming that, layers are likely registered according to the same order as they are invoked in the forward pass. Hence, the reverse order should approximately represent the gradient computation order in the backward pass. Admittedly, this is not a perfect solution, but is an approximation that we can rely on with minimum engineering overhead.

Second, it is possible that one training iteration only involves a sub-graph in the model and the sub-graph can be different from iteration to iteration, meaning that some gradients might be skipped in some iterations. However, as gradient-to-bucket mapping is determined at the construction time, those absent gradients would leave some buckets never seeing the final autograd hook and failing to mark the bucket as ready. As a result, the backward pass could hang. 
Fig.~\ref{fig:caveats}~(b) shows an example, where the parameter corresponding to gradient $g_3$ is skipped in one iteration, leading to the absent of the ready signal for $g_3$.
To address this problem, \code{DDP} traverses the autograd graph from the output tensors of the forward pass to find all participating parameters. The readiness of those participating tensors is a sufficient signal to conclude the completion of the backward pass. Therefore, \code{DDP} can avoid waiting for the rest of the parameter gradients by proactively marking them ready at the end of the forward pass. Note that, this change does not prevent us from developing non-intrusive APIs, because application directly invokes the \code{forward} function on \code{DDP} and hence \code{DDP} can easily insert this step in its member function. 

\vspace{-0.8em}

\begin{algorithm}
\footnotesize
\caption{DistributedDataParallel}\label{algo:reduction}
\KwIn{Process rank $r$, bucket size cap $c$, local model $net$}
\SetKwFunction{constructor}{constructor}
\SetKwFunction{forward}{forward}
\SetKwFunction{hook}{autograd\_hook}
\SetKwProg{Fn}{Function}{:}{}
\Fn{\constructor{net}}{
\If{r=0}{
broadcast $net$ states to other processes\\
}
init buckets, allocate parameters to buckets in the reverse order of net.\text{parameters()}\\
\For{p \textbf{in} net.parameters()}{
    acc $\leftarrow p.\text{grad\_accumulator}$\\
    acc $\rightarrow \text{add\_post\_hook}(\text{autograd\_hook})$\\
}
}
\Fn{\forward{inp}}{
    out = $net$(inp)\\
    traverse autograd graph from out and mark unused parameters as ready\\
    \text{\textbf{return}} out
}
\Fn{\hook{param\_index}}{
get bucket $b_i$ and bucket \emph{offset} using param\_index\\
get parameter $var$ using \emph{param\_index}\\
view $\leftarrow b_i.narrow($\emph{offset}, $var.size())$\\
view.copy\_(var.grad)\\
\If{all grads in $b_i$ are ready}{
    mark $b_i$ as ready\\
}
launch AllReduce on ready buckets in order\\
\If{all buckets are ready}{
    block waiting for all AllReduce ops\\
}
}
\end{algorithm}

\vspace{-0.8em}

\begin{figure}
\centering
  \includegraphics[width=0.9\linewidth]{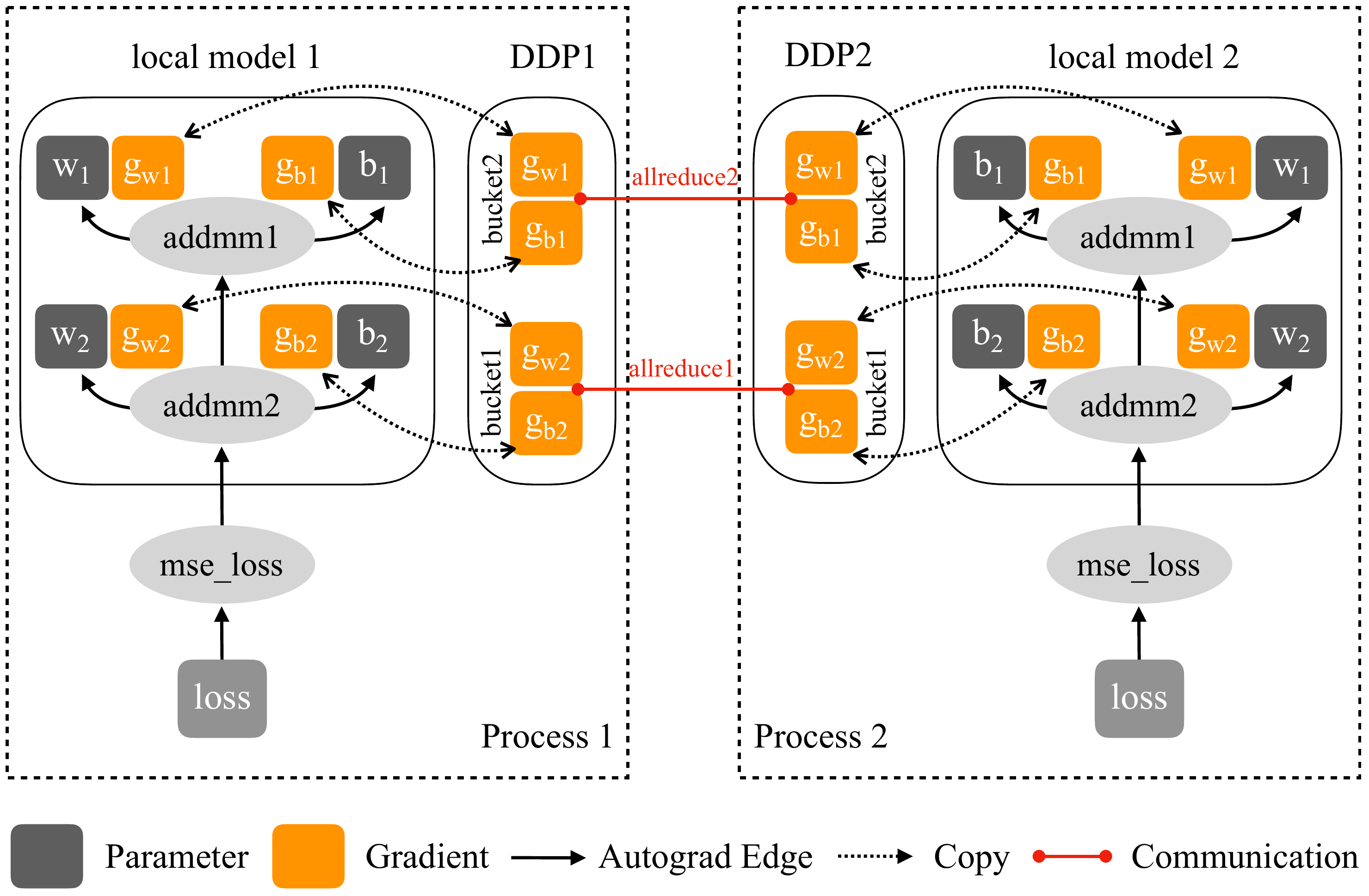}
  \vspace{-1em}
  \caption{Distributed Gradient Reduction}\label{fig:ddp}
  \vspace{-1.5em}
\end{figure}

Algorithm~\ref{algo:reduction} presents the pseudo-code of \code{DDP}. The constructor contains two major steps, broadcasting model states and installing autograd hooks. \code{DDP}'s \code{forward} function is a simple wrapper of the local model's \code{forward}, and traverses the autograd graph to mark unused parameters at the end. The \code{autograd\_hook} takes the internal parameter index as input, which helps to find the parameter tensor and its belonging bucket. It writes the local gradient to the correct offset in the bucket and then launches the asynchronous \code{AllReduce} operation. There is an additional finalizing step omitted in the pseudo-code that waits for \code{AllReduce} operations and writes the value back to gradients at the end of the backward pass.  Fig.~\ref{fig:ddp} elucidates how \code{DDP} interacts with the local model during the forward and backward passes.

The above solution works for most use cases. However, as \code{DDP} always computes the average of all gradients and writes them back to parameter \code{.grad} field, an optimizer cannot distinguish whether a gradient has participated in the last backward pass or not. Due to the decoupled design of \code{DDP} and the optimizer, there is no side channel for \code{DDP} to allude that information to the optimizer. Without this information, the training process could suffer from regressions on model accuracy, \emph{e.g.}, when the optimizer uses gradient absence information to skip updating momentum values. To tackle this problem, \code{DDP} should only touch gradients that are indeed involved in the backward pass. Nevertheless, this information cannot be extracted from the local autograd graph alone, because locally absent gradients might still be involved in the forward/backward pass in a peer \code{DDP} process. Therefore, \code{DDP} uses a bitmap to keep track of local parameter participants and launches one additional \code{AllReduce} to collect globally unused parameters. Unfortunately, \code{DDP} cannot coalesce this bitmap into other gradient \code{AllReduce} operations due to the potential mismatch in element types. Such additional overhead only materializes when the application explicitly tells \code{DDP} to look for unused parameters, and hence the price is only paid when necessary.

\subsubsection{Gradient Accumulation}\label{sec:accumulation}

One common technique to speed up distributed data parallel training is to reduce gradient synchronization frequencies. Instead of launching \code{AllReduce} in every iteration, the application can conduct $n$ local training iterations before synchronizing gradients globally. This is also helpful if the input batch is too large to fit into a device, where the application could split one input batch into multiple micro-batches, run local forward and backward passes on every micro-batch, and only launch gradient synchronization at the boundaries of large batches. Theoretically, this should produce the same results as if all data in the large batch is processed in one shot, as gradients will simply be accumulated to the same tensor. However, this conflicts with the gradient reduction algorithm discussed in Section~\ref{sec:overlap} to some degree. That algorithm would mark unused parameters as ready at the end of every forward pass, while those unused parameters in one iteration still could participate in subsequent iterations. Moreover, \code{DDP} cannot distinguish whether the application plans to immediately invoke \code{optimizer.step()} after backward or accumulate gradients through multiple iterations. Therefore, we need to introduce one additional interface (\emph{i.e.}, \code{no\_sync}) for this use case. Below is an example code snippet.
\begin{lstlisting}[language=Python]
ddp = DistributedDataParallel(net)
with ddp.no_sync():
  for inp, exp in zip(inputs, expected_outputs):
    # no synchronization, accumulate grads
    loss_fn(ddp(inp), exp).backward()  
# synchronize grads
loss_fn(ddp(another_inp), another_exp).backward()  
opt.step()
\end{lstlisting}
Under the hood, the implementation for \code{no\_sync} is very simple. The context manager just toggles a flag on entering and exiting the context, and the flag is consumed in the \code{forward} function of \code{DDP}. In \code{no\_sync} mode, all \code{DDP} hooks are disabled, and the first backward pass out of the context will synchronize the accumulated gradients altogether. The information of globally unused parameters also accumulates in the bitmap, and serves when the next communication takes place.
\subsection{Collective Communication}\label{sec:collective}

Distributed data parallel training uses a special communication pattern, where every participant provides an equally-sized tensor and collects the global sum across all participants. This can certainly be implemented as a gather operator followed by local reductions on every participant using point-to-point communication, but that would forfeit opportunities for performance optimizations~\cite{nccl:tree_reduce}. \code{DDP} is built on top of collective communication libraries, including three options, NCCL~\cite{nccl}, Gloo~\cite{gloo}, and MPI~\cite{mpi}. \footnote{Please refer to documents of the three libraries for their design and implementation.} \code{DDP} takes the APIs from the three libraries and wraps them into the same \code{ProcessGroup} API. The name heralds that \code{ProcessGroup} expects multiple processes to work collectively as a group. All \code{ProcessGroup} instances construct at the same time, which is implemented using a rendezvous service, where the first arrival will block waiting until the last instance joins. For NCCL backend, the ProcessGroup maintains a dedicated set of CUDA streams for communication, so that it will not block the computation in the default stream. As all communications are collective operations, subsequent operations on all \code{ProcessGroup} instances must match in size and type and follow the same order. Using the same \code{ProcessGroup} API for all libraries allows us to experiment with different communication algorithms with the same \code{DDP} implementation. For example, PyTorch v1.5 provides a composite round-robin \code{ProcessGroup} implementation, which takes a list of \code{ProcessGroup} instances and dispatches collective communications to those \code{ProcessGroup} instances in a round-robin manner. By using round-robin \code{ProcessGroup}s, \code{DDP} can attain higher bandwidth utilization if a single NCCL, Gloo, or MPI \code{ProcessGroup} is unable to saturate the link capacity.

\section{Implementation}\label{sec:impl}

The implementation of \code{DDP} has evolved several times in the past few releases. This section focus on the current status as of PyTorch v1.5.0. \code{DDP} implementation lives both in Python and C++ files, with Python exposing the API and composing non-performance-critical components, and C++ serving the core gradient reduction algorithm. The Python API calls into C++ core through Pybind11~\cite{pybind}.

\subsection{Python Front-end}

The \code{DDP} \code{nn.module} is implemented in \code{distributed.py}, which contains user-facing components, including the constructor, the \code{forward} function, and the \code{no\_sync} context manager. Besides the general ideas highlighted in Section~\ref{sec:design}, there are several implementation details in the Python front-end that shapes the behavior of \code{DDP}.


\textbf{Configuable Knobs} are exposed in the \code{DDP} constructor API, including 1) \code{process\_group} to specify a process group instance for \code{DDP} to run \code{AllReduce}, which helps to avoid messing up with the default process group, 2) \code{bucket\_cap\_mb} to control the \code{AllReduce} bucket size, where applications should tune this knob to optimize training speed, and 3) \code{find\_unused\_parameters} to toggle whether \code{DDP} should detect unused parameters by traversing the autograd graph.
\textbf{Model Device Affinity} in the local model also governs \code{DDP}'s behavior, especially if the model spans multiple devices, which is common when the model is too large to fit into a single device. For large models, applications can place different layers of the model onto difference devices, and use \code{Tensor.to(device)} API to move intermediate output from one device to another. \code{DDP} also works with multi-device models. As long as the \code{device\_ids} argument is \code{None} or an empty list, \code{DDP} will inspect the model, perform sanity checks and apply configurations accordingly. Then, it treats the multi-device model as one entirety. 

\textbf{Model Buffers} are necessary when layers (\emph{e.g.}, \code{BatchNorm}) need to keep track of states like the running variance and the running mean. \code{DDP} supports model buffers by letting the process with the rank 0 to take the authority. If the model contains buffers, \code{DDP} will broadcast the buffer values from rank 0 process to all other processes before starting the forward pass on the local model. This behavior is also compatible with the \code{no\_sync} mode. When \code{no\_sync} mode is enabled, it sets a flag in the forward pass properly to indicate whether it expects gradient reductions in the immediate backward pass. If the communication takes place, \code{DDP} will then broadcast buffers prior to the subsequent forward pass. 


\subsection{Core Gradient Reduction}

Major development efforts are spent in gradient reduction as it is the most performance-critical step in \code{DDP}. The implementation lives in \code{reducer.cpp} which consists of four main components, namely, building parameter-to-bucket map, installing autograd hooks, launching bucket \code{AllReduce}, and detecting globally unused parameters. This section expatiates on these four components.

\textbf{Parameter-to-Bucket Mapping} has considerable impact on \code{DDP} speed. In every backward pass, tensors are copied from all parameter gradients to buckets, and averaged gradients are copied back after \code{AllReduce}. To accelerate copy operations, buckets are always created on the same device as the parameters. If the model spans multiple devices, \code{DDP} takes device affinity into consideration to make sure that all parameters in the same bucket are on the same device. The order of \code{AllReduce} also makes a difference, as it dictates how much communication can overlap with computation. \code{DDP} launches \code{AllReduce} in the reverse order of \code{model.parameters()}.

\textbf{Autograd Hook} is the entry point for \code{DDP} in the backward pass. During construction, \code{DDP} loops over all parameters in the model, finds the gradient accumulator on every parameter, and installs the same post-hook function to every gradient accumulator. The gradient accumulator will fire post hooks when the corresponding gradient is ready, and \code{DDP} will figure out when an entire bucket is ready  to launch an \code{AllReduce} operation. However, as there is no guarantee on the order of gradient readiness, \code{DDP} cannot selectively pick parameters to install hooks. In the current implementation, each bucket keeps a count of pending gradients. Each post-hook function decrements the count, and \code{DDP} marks a bucket as ready when that count reaches zero. In the next \code{forward} pass, \code{DDP} replenishes the pending gradient count for every bucket.

\textbf{Bucket Allreduce} is the main source of communication overhead in \code{DDP}. On one hand, packing more gradients into the same bucket would reduce the amortized system overhead of communication. One the other hand, using a large bucket size would result in longer lead time for reduction, as each bucket needs to wait for more gradients. Hence, bucket size is the key trade-off. By default, each bucket is \code{25MB} in size. Applications should measure their impact empirically and set it to the optimal value for their use cases.

\textbf{Globally Unused Parameters}' gradients should stay intact during the forward and the backward passes. Detecting unused parameters requires global information, as one parameter could be absent in one \code{DDP} process during one iteration, but participates training in the same iteration in another process. \code{DDP} maintains local unused parameter information in a bitmap, and launches an additional \code{AllReduce} to gather a global bitmap. As the bitmap is much smaller than tensor sizes, instead of creating per-bucket bitmaps, all parameters in the model share the same bitmap. The bitmap lives on CPU to avoid launching dedicated CUDA kernels for each update. However, some \code{ProcessGroup} backends might not be able to run \code{AllReduce} on CPU tensors. For example, \code{ProcessGroupNCCL} only supports CUDA tensors. Moreover, as \code{DDP} should work with any custom \code{ProcessGroup} backend, it cannot make assumptions that all backends support CPU tensors. To address this problem, \code{DDP} maintains another bitmap on the same device as the first model parameter, and invokes a non-blocking copy to move the CPU bitmap to the device bitmap for collective communications. 

\section{Evaluation}\label{sec:eval}


\begin{figure}
\centering
  \includegraphics[width=0.7\linewidth]{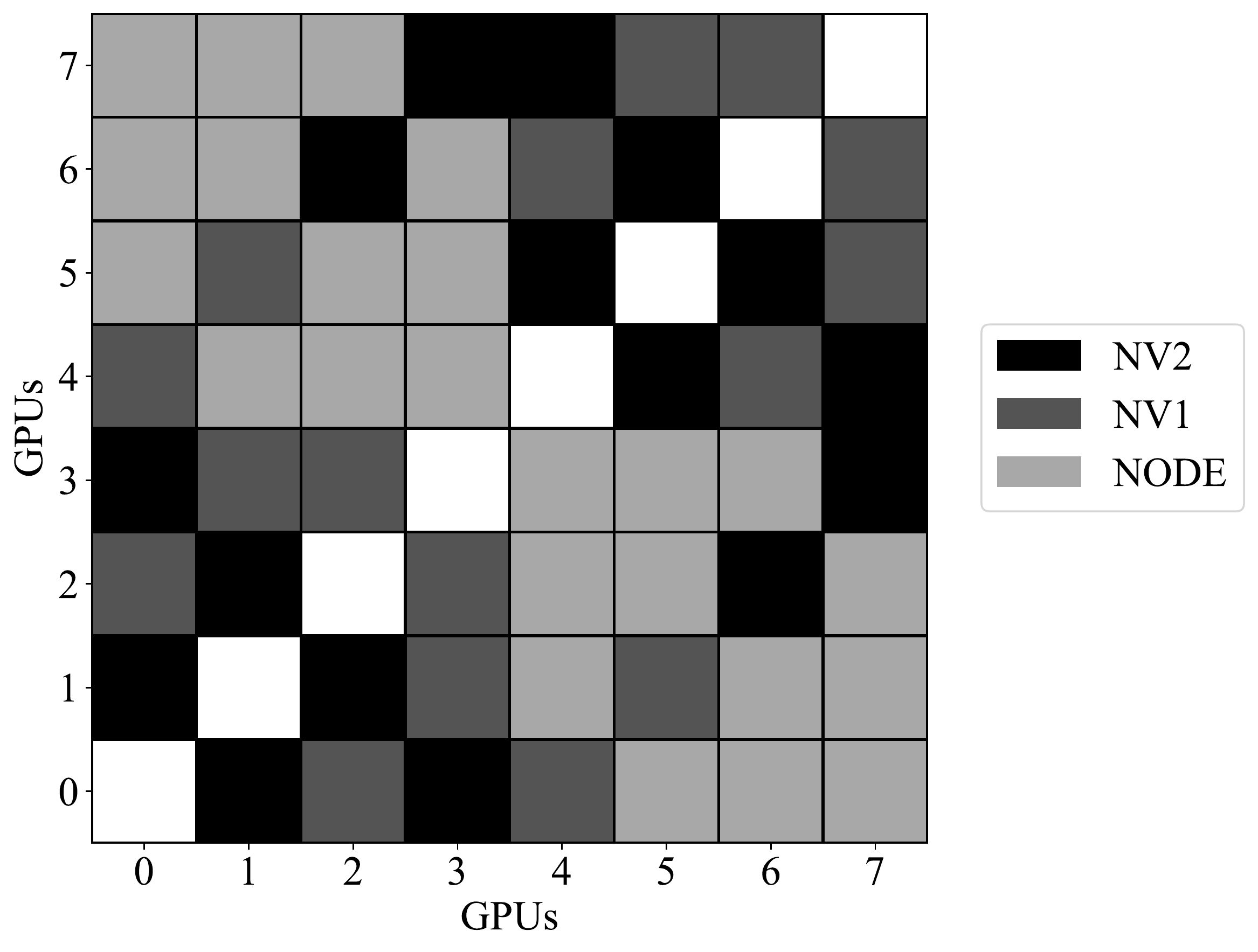}
  \vspace{-1em}
  \caption{GPU Connection Topology}\label{fig:topo}
  \vspace{-1.5em}
\end{figure}

This section presents the evaluation results of PyTorch \code{DDP} using an exclusive 32 GPU cluster and a shared entitlement. In the exclusive cluster, the GPUs are located on 4 servers, connected using Mellanox MT27700 ConnectX-4 100GB/s NIC. All 4 servers reside in the same rack, and each server is equipped with 8 NVIDIA Tesla V100 GPUs. Fig.~\ref{fig:topo} shows the interconnection of the 8 GPUs within the same server. We only use the shared entitlement when a set of experiments require more than 32 GPUs.
In the shared entitlement, we submit jobs to run on different numbers of GPUs where different jobs can run on different machines, and hence the hardware and network connectivity can vary from job to job. Although the disparity in the test environment can lead to different latency measures even for the same code, we pack the same set of experiments into the same job, so that the trend shown in the same curve is still meaningful.


We measure \code{DDP} per iteration latency and scalability using two popular models, ResNet50~\cite{resnet} and BERT~\cite{bert}, to represent typical vision and NLP applications. Most experiments use randomly generated synthetic inputs and labels, which are sufficient as the purpose is to compare per iteration latency instead of model accuracy. Experiments compute losses using the \code{CrossEntropyLoss} function and update parameters using the \code{SGD} optimizer. Configurations for accuracy-related experiments will be explained in detail close to their presentations.

\begin{figure}
\centering
  \includegraphics[width=0.8\linewidth]{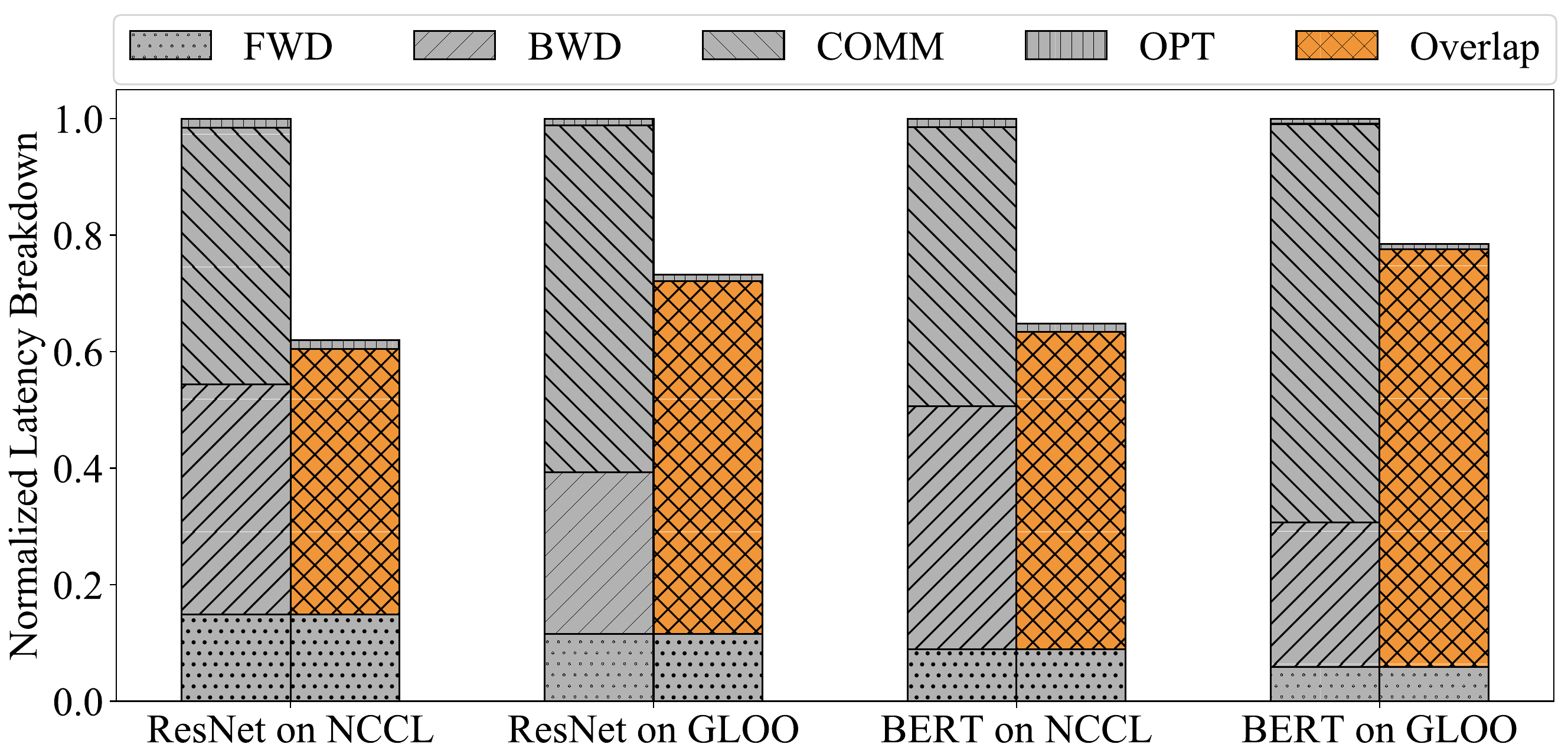}
  \vspace{-1em}
  \caption{Per Iteration Latency Breakdown}\label{fig:breakdown}
  \vspace{-1.5em}
\end{figure}

\begin{figure*}[!htb]
\begin{minipage}[c]{0.245\textwidth}
  \centering
  \includegraphics[width=\linewidth]{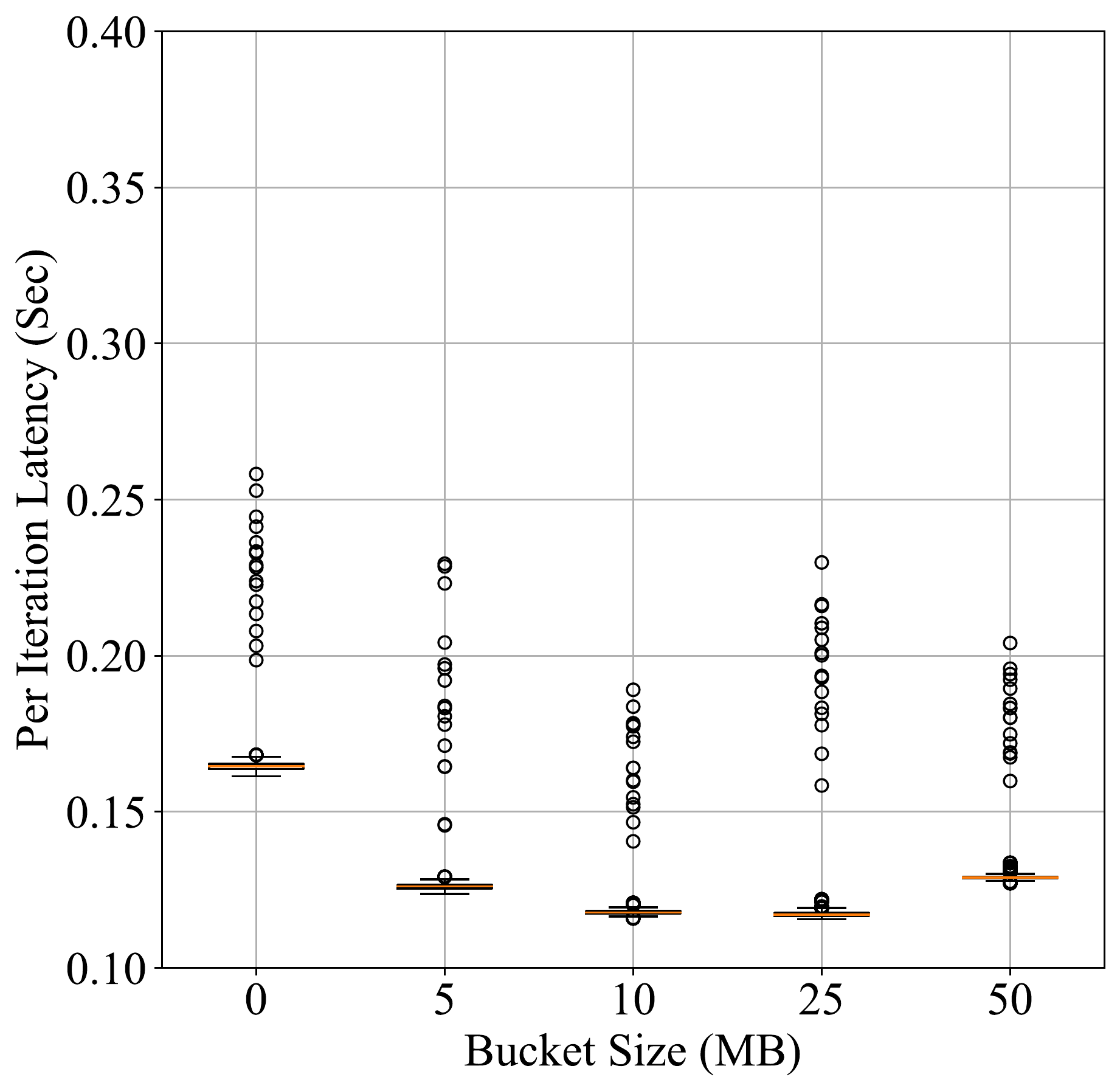}
  (a) ResNet50 on NCCL
\end{minipage}
\begin{minipage}[c]{0.245\textwidth}
  \centering
  \includegraphics[width=\linewidth]{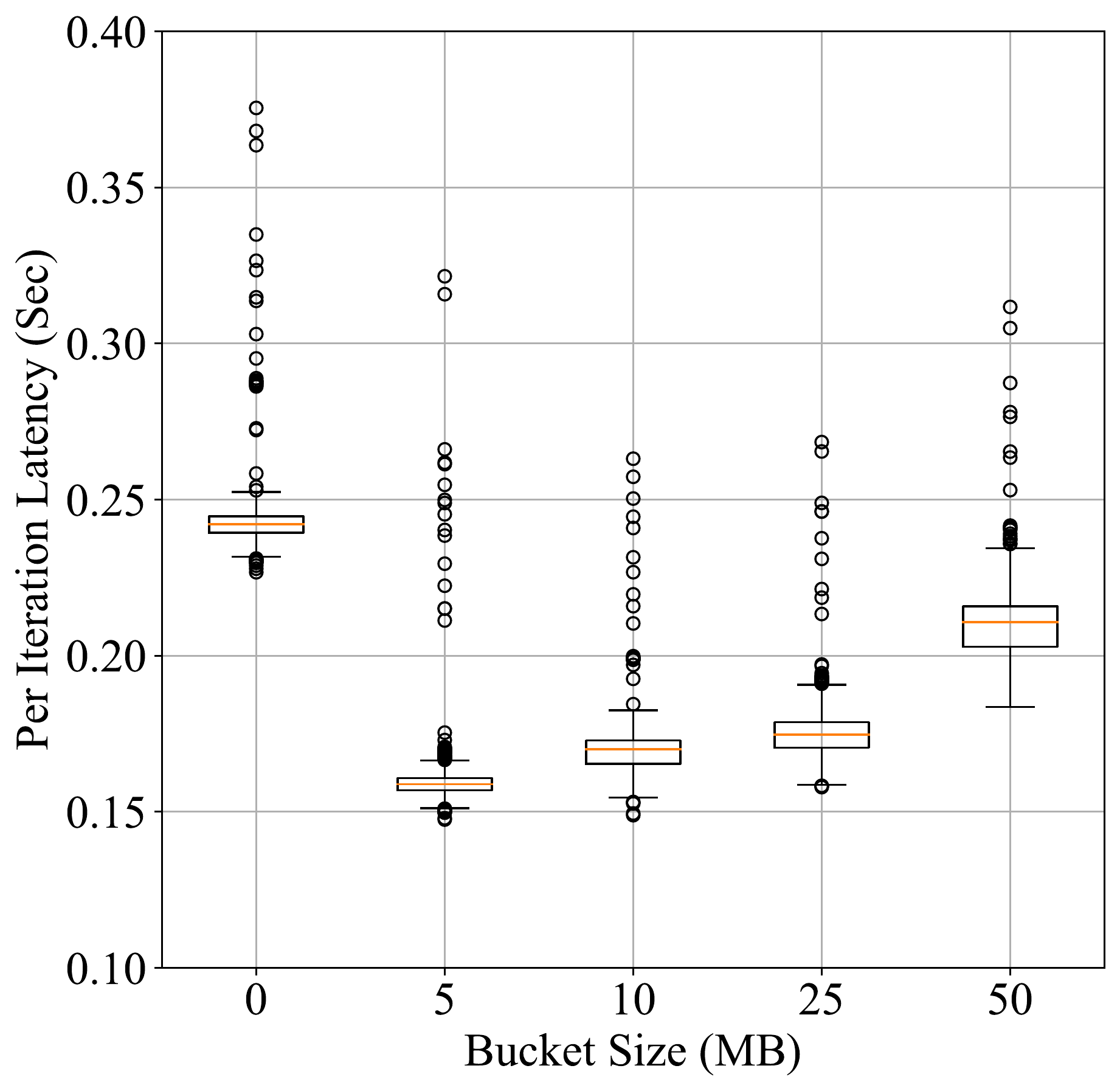}
  (b) ResNet50 on Gloo
\end{minipage}
\begin{minipage}[c]{0.245\textwidth}
  \centering
  \includegraphics[width=\linewidth]{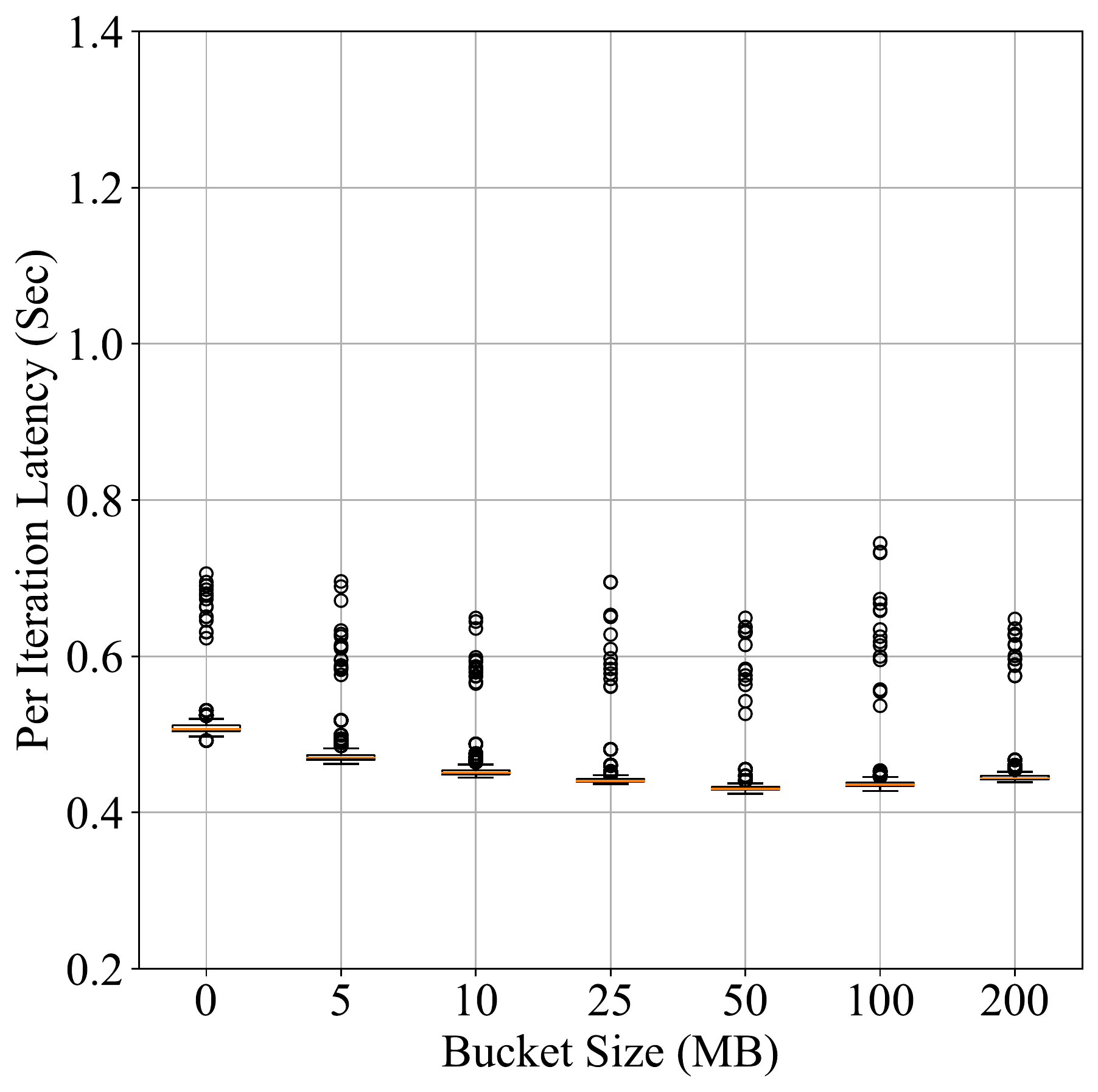}
  (c) BERT on NCCL
\end{minipage}
\begin{minipage}[c]{0.245\textwidth}
  \centering
  \includegraphics[width=\linewidth]{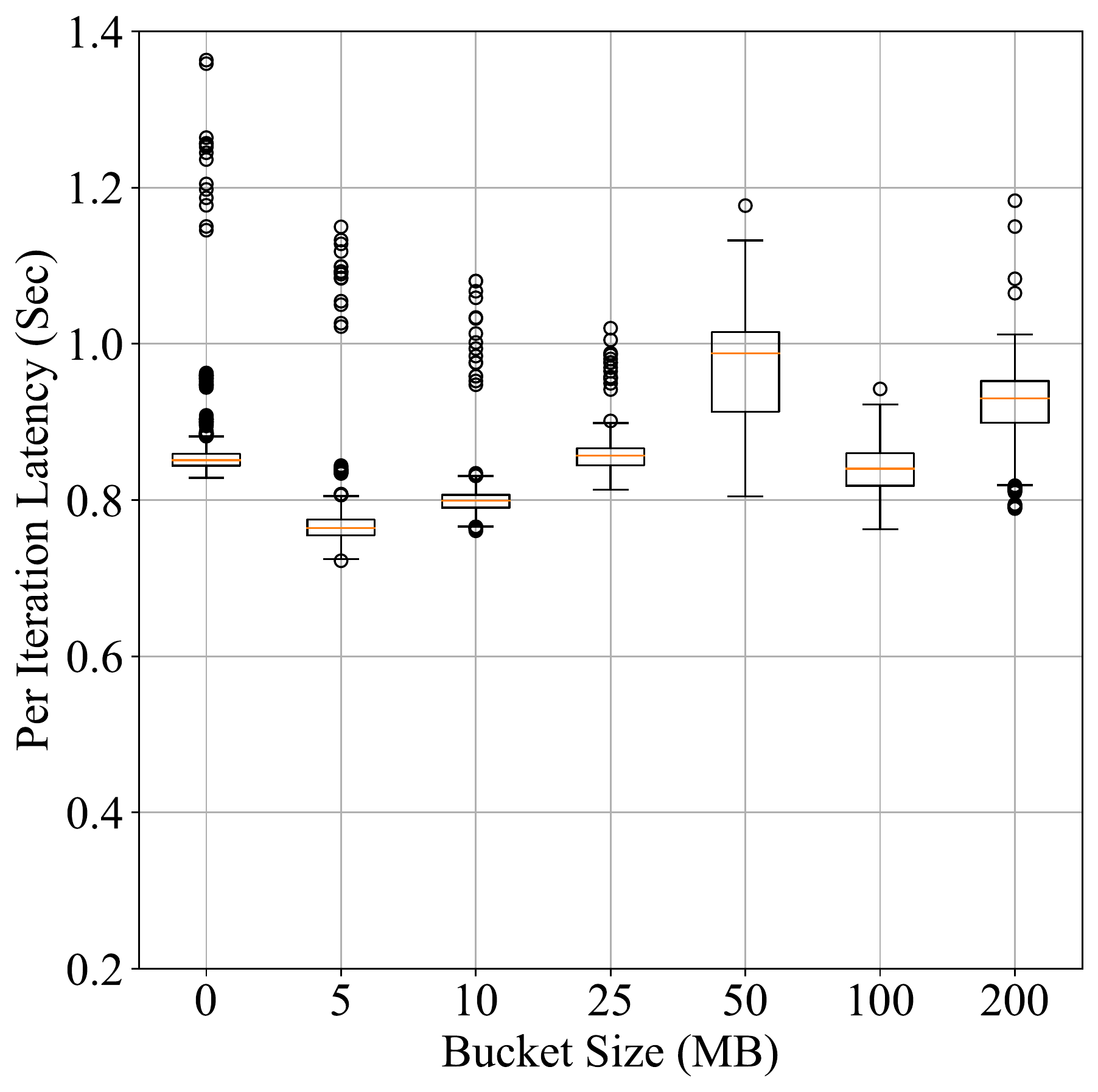}
  (d) BERT on Gloo
\end{minipage}
\vspace{-1em}
\caption{Per Iteration Latency vs Bucket Size on 16 GPUs}\label{fig:bucket_size}
\vspace{-0.8em}
\end{figure*}

\begin{figure*}[!htb]
\begin{minipage}[c]{0.245\textwidth}
  \centering
  \includegraphics[width=\linewidth]{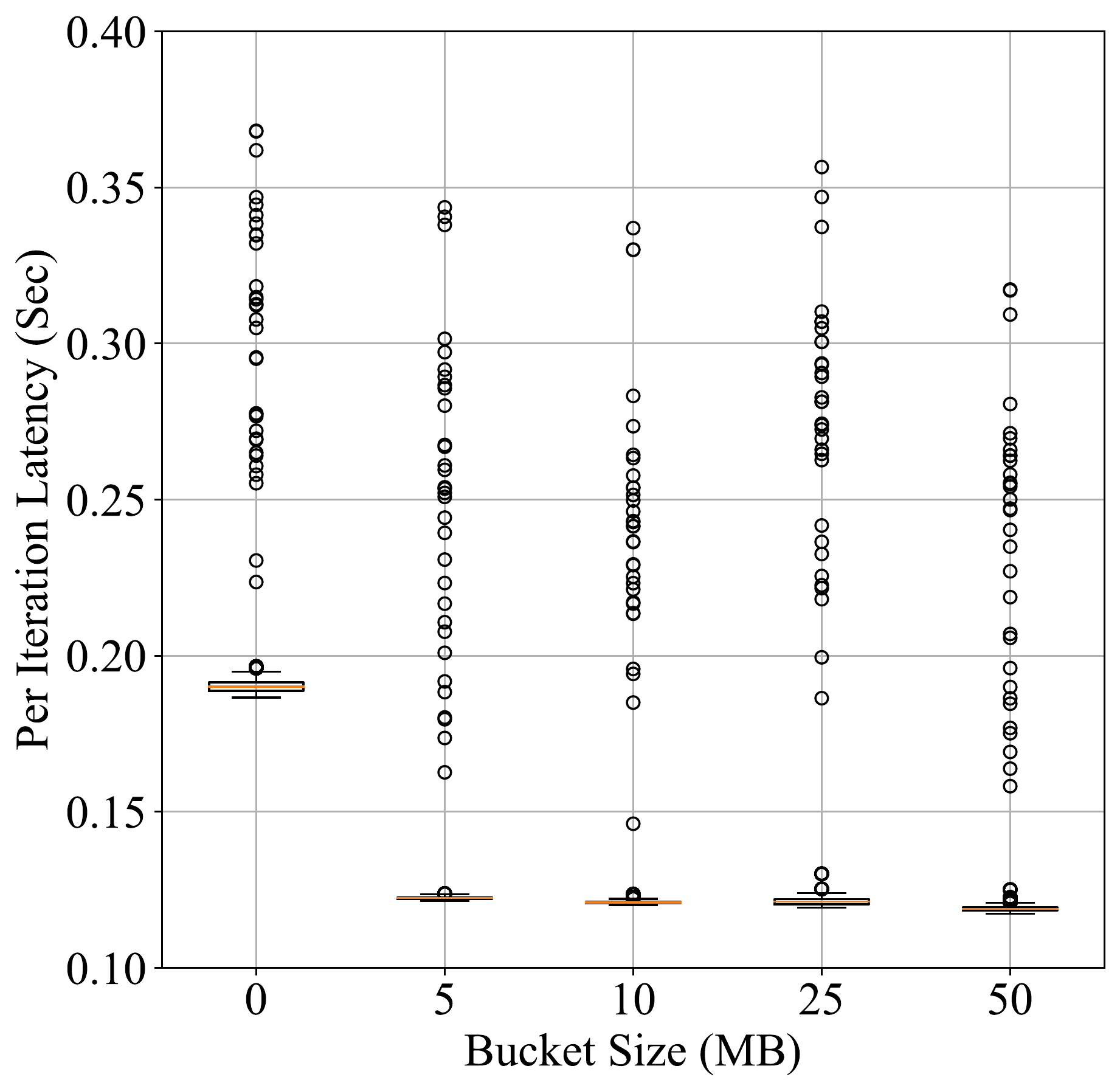}
  (a) ResNet50 on NCCL
\end{minipage}
\begin{minipage}[c]{0.245\textwidth}
  \centering
  \includegraphics[width=\linewidth]{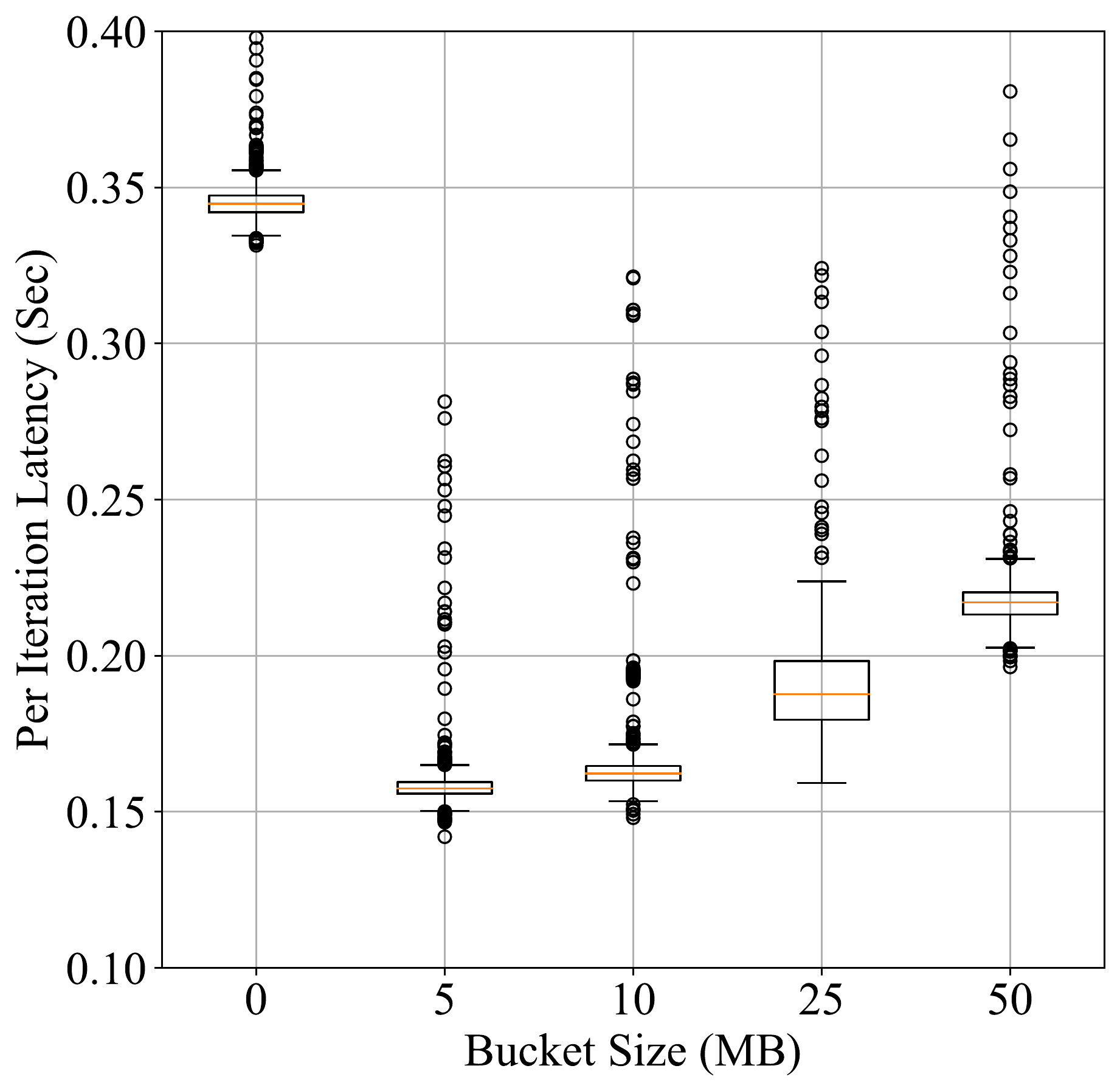}
  (b) ResNet50 on Gloo
\end{minipage}
\begin{minipage}[c]{0.245\textwidth}
  \centering
  \includegraphics[width=\linewidth]{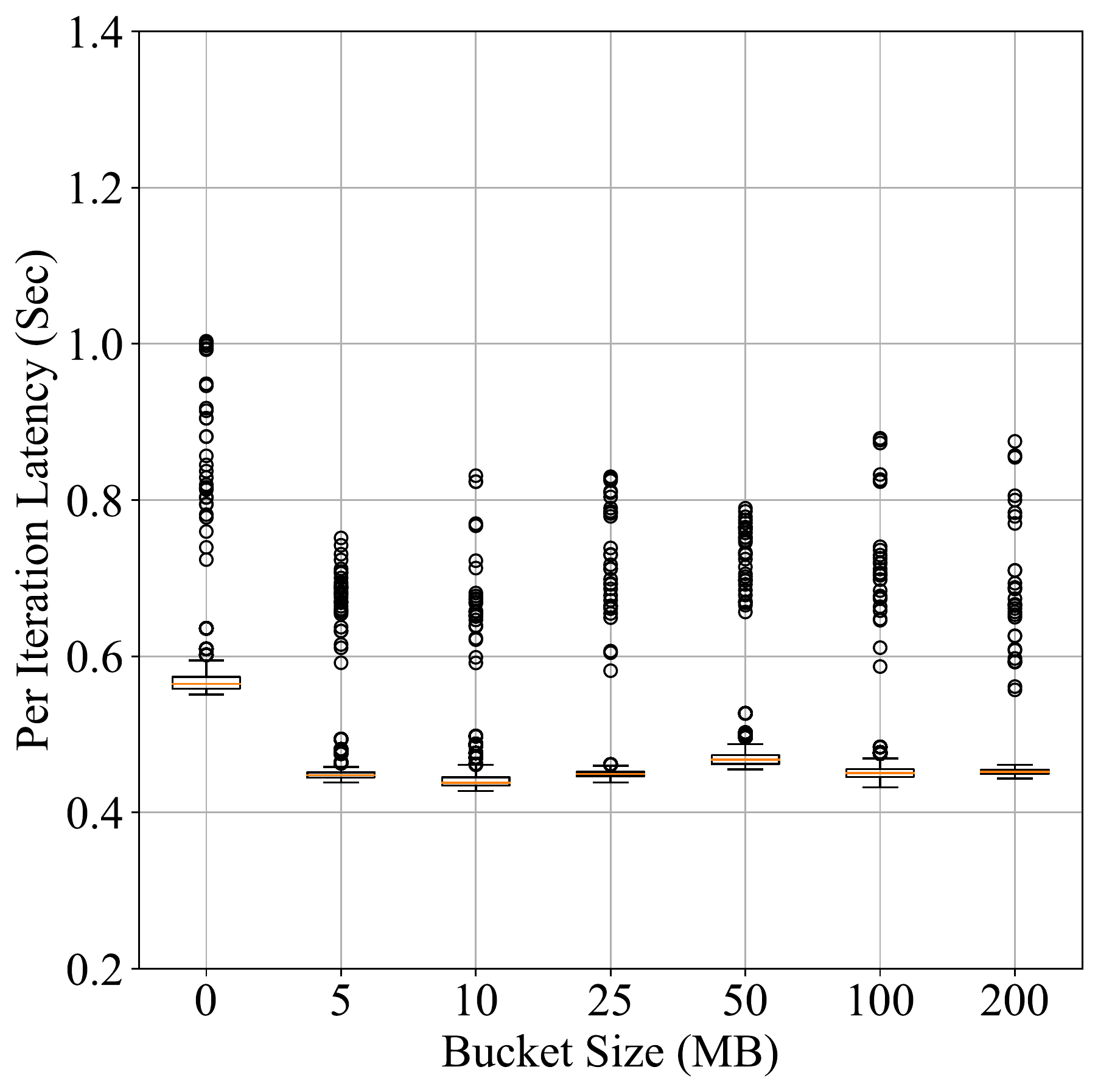}
  (c) BERT on NCCL
\end{minipage}
\begin{minipage}[c]{0.245\textwidth}
  \centering
  \includegraphics[width=\linewidth]{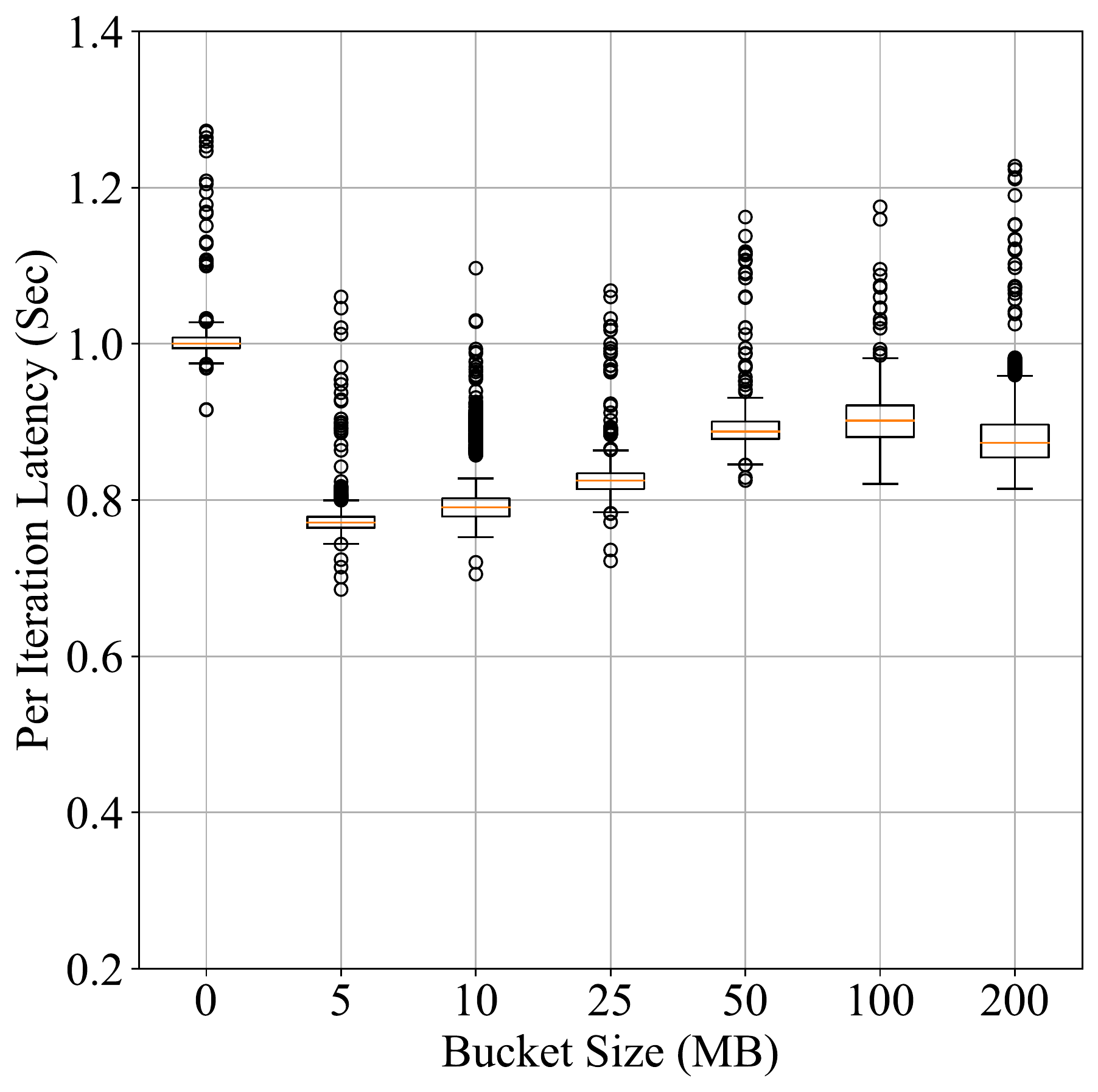}
  (d) BERT on Gloo
\end{minipage}
\vspace{-1em}
\caption{Per Iteration Latency vs Bucket Size on 32 GPUs}\label{fig:bucket_size_32}
\vspace{-1em}
\end{figure*}

\subsection{Latency Breakdown}

A typical training iteration contains three steps: forward pass to compute loss, backward pass to compute gradients, and optimizer step to update parameters. 
With \code{DDP}, the backward pass involves local computation and \code{AllReduce} communication. To demonstrate the effectiveness of overlapping computation with communication, Fig.~\ref{fig:breakdown} plots the latency breakdown when using NCCL and Gloo backends for ResNet50 and BERT models respectively. All experiments are conducted using 32 GPUs across 4 machines. To visually compare the speedup on different model and backend combinations, we normalize the total latency to 1 for all non-overlapping cases. 
The results demonstrate that the backward pass is the most time-consuming step with PyTorch \code{DDP} training, as \code{AllReduce} communications (\emph{i.e.}, gradient synchronization) are completed in this step. This observation justifies that the \code{DDP} backward pass deserves the most efforts for improvements. Within the backward pass, the communication step takes more than half of the total delay and this is exacerbated with the increase of model size. Between these two backends, NCCL is considerably faster than GLOO. The speedup is most effective when the computation and communication take roughly the same amount of time as they can overlap more. The overlapping approach helps ResNet and BERT on NCCL attain 38.0\% and 35.2\% speedup. With GLOO backend, the gain shrinks to 26.8\% and 21.5\% respectively, as GLOO communication becomes the dominating delay in the backward pass.  

\subsection{Bucket Size}

To avoid launching an excessive number of \code{AllReduce} operations, \code{DDP} organizes small gradients into larger buckets and synchronizes each bucket using an \code{AllReduce} operation. With this design, bucket size is an important configuration knob. \code{DDP} exposes this knob to applications through \code{bucket\_cap\_mb} argument. No single bucket size can best serve all applications. This value should be measured and determined empirically. The default value of \code{bucket\_cap\_mb} is 25MB, which is our best effort estimation based experiences. The following experiments also confirm this is a reasonable choice for ResNet50 and BERT. This section compares per iteration latency across different bucket sizes using 16 GPUs on two machines. Zero bucket size means each gradient will be communicated on its own as soon as it is ready. This serves as a baseline on one extreme of the bucket size spectrum. The other extreme is communication all gradients in one short, which is skipped as results in Fig.~\ref{fig:bucket_size} and Fig.~\ref{fig:bucket_size_32} clearly show the best option for both ResNet50 and BERT is somewhere in the middle.

Fig.~\ref{fig:bucket_size}~(a) uses box-whisker to illustrate how bucket size affects per iteration latency on ResNet50 with NCCL backend. The x-axis is the bucket size in MBs, and Y-axis per iteration latency in seconds. The outliers are the tiny delay spikes at 100 iteration boundaries caused by \code{DDP} instance re-construction and input data regeneration. Other than that, delays of most iterations concentrate in a very narrow time range, which also agrees with the results shown in Fig.~\ref{fig:breakdown}~(a). The results show that the highest speed is achieved between 10MB and 25MB bucket sizes. Fig.~\ref{fig:bucket_size}~(b) presents the same measurements for Gloo backend. The results are different from NCCL backend in two ways, 1) per iteration latency falls into a large range, 2) the 5MB bucket size attains higher speed compared to 10MB and 25MB. The first difference matches with Fig.~\ref{fig:breakdown}~(b). To understand the second difference, let us revisit Fig.~\ref{fig:allreduce}~(b) on Gloo \code{AllReduce} latency across different tensor sizes. It's clear that the total \code{AllReduce} time fluctuates around the same level when the bucket size is larger than 512KB. Therefore, larger bucket sizes beyond 512KB with Gloo backend would only mean longer waiting time for gradients, which leads to longer per iteration latency. Fig.~\ref{fig:bucket_size}~(c) and (d) show the measurements for BERT model. As BERT model contains 15X more parameters compared to ResNet50, intuitively, it should benefit from larger buckets as larger communication overheads would dwarf the waiting time for the first bucket. The results verified the intuition with NCCL backend, where 50MB bucket size leads to the best performance. However, with Gloo backend, 5MB bucket size still wins with the lowest per iteration latency.

Fig.~\ref{fig:bucket_size_32} presents the results of the same set of experiments but on 32 GPUs. In this case, the outliers span a larger range, which is not surprising as synchronizations usually take longer with more participants and the impact of strangler is more prominent. Fig.~\ref{fig:bucket_size_32}~(a) and (b) both suggest that 0MB bucket size leads to obviously longer per iteration latency on 32 GPUs compared to 16 GPUs, as per-gradient reductions on a larger cluster are expected to be slower. However, when bucket size is set to above 5MB, scaling from 16 GPUs to 32 GPUs does not lead to a noticeable speed regression. This is probably because although individual \code{AllReduce} operations is expected to be slower, asynchronous execution and parallelism could help to hide the overall delay.

\subsection{Scalability}

\begin{figure*}[!htb]
\begin{minipage}[c]{0.245\textwidth}
  \centering
  \includegraphics[width=\linewidth]{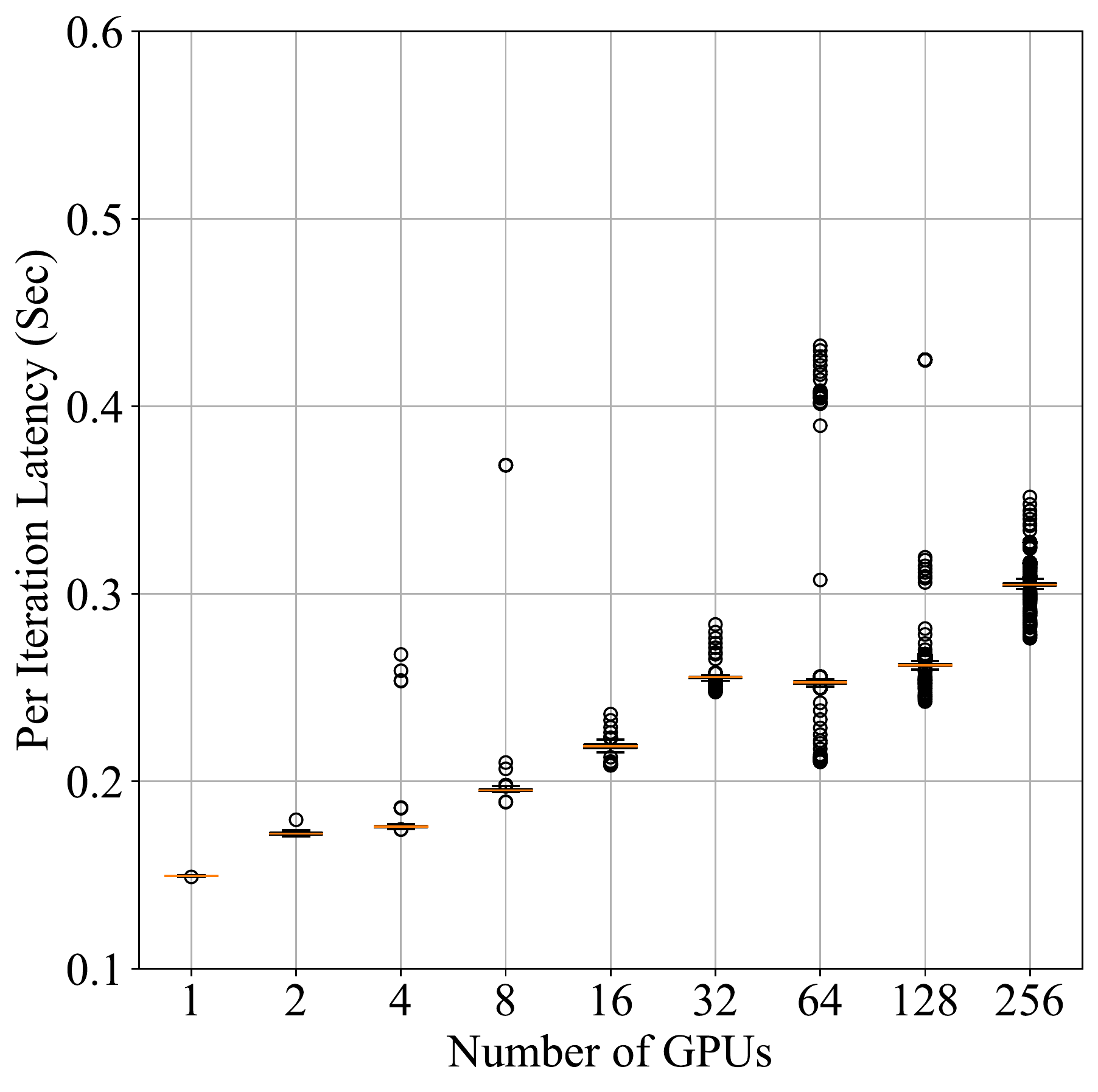}
  (a) ResNet50 on NCCL
\end{minipage}
\begin{minipage}[c]{0.245\textwidth}
  \centering
  \includegraphics[width=\linewidth]{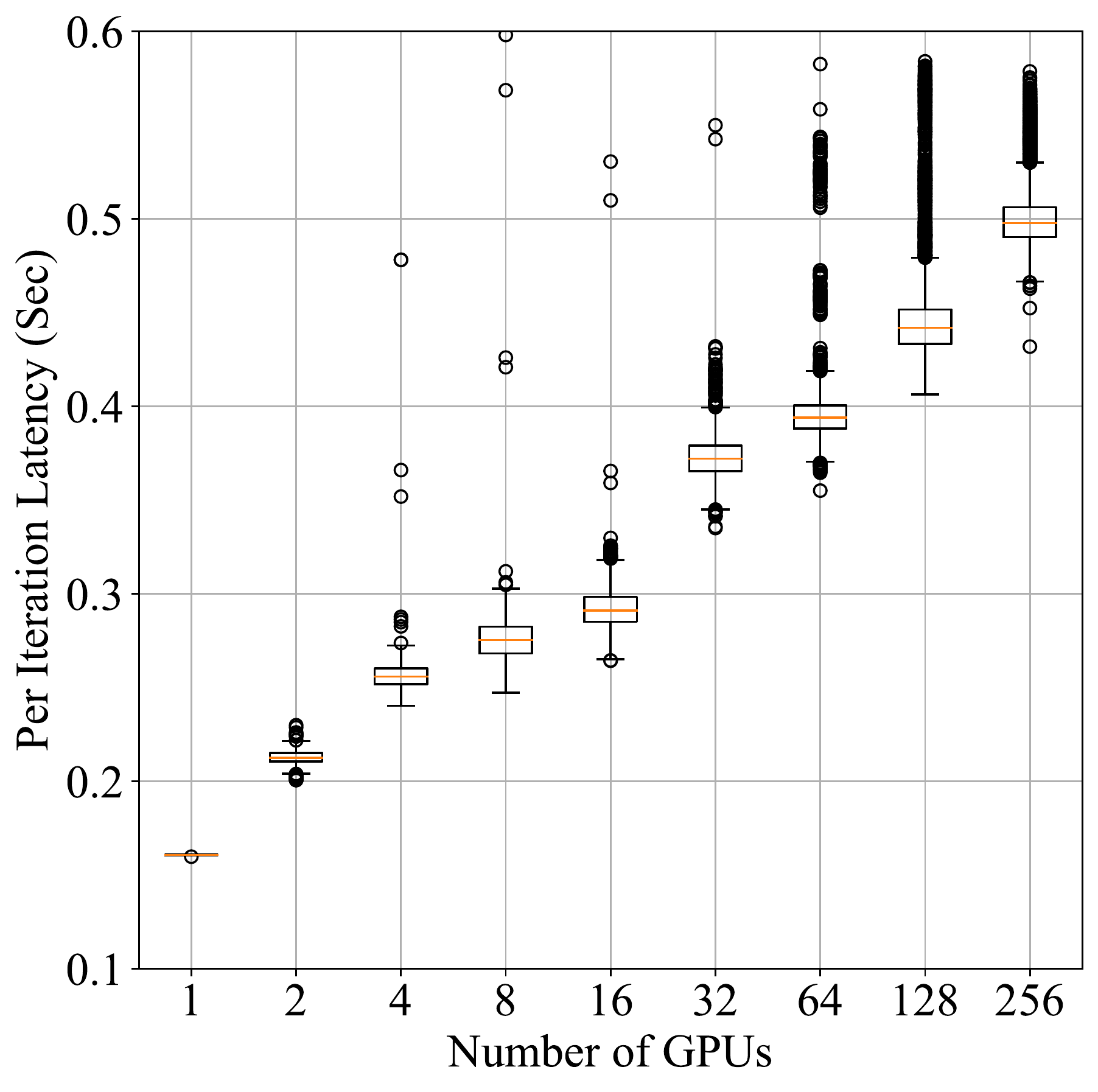}
  (b) ResNet50 on Gloo
\end{minipage}
\begin{minipage}[c]{0.245\textwidth}
  \centering
  \includegraphics[width=\linewidth]{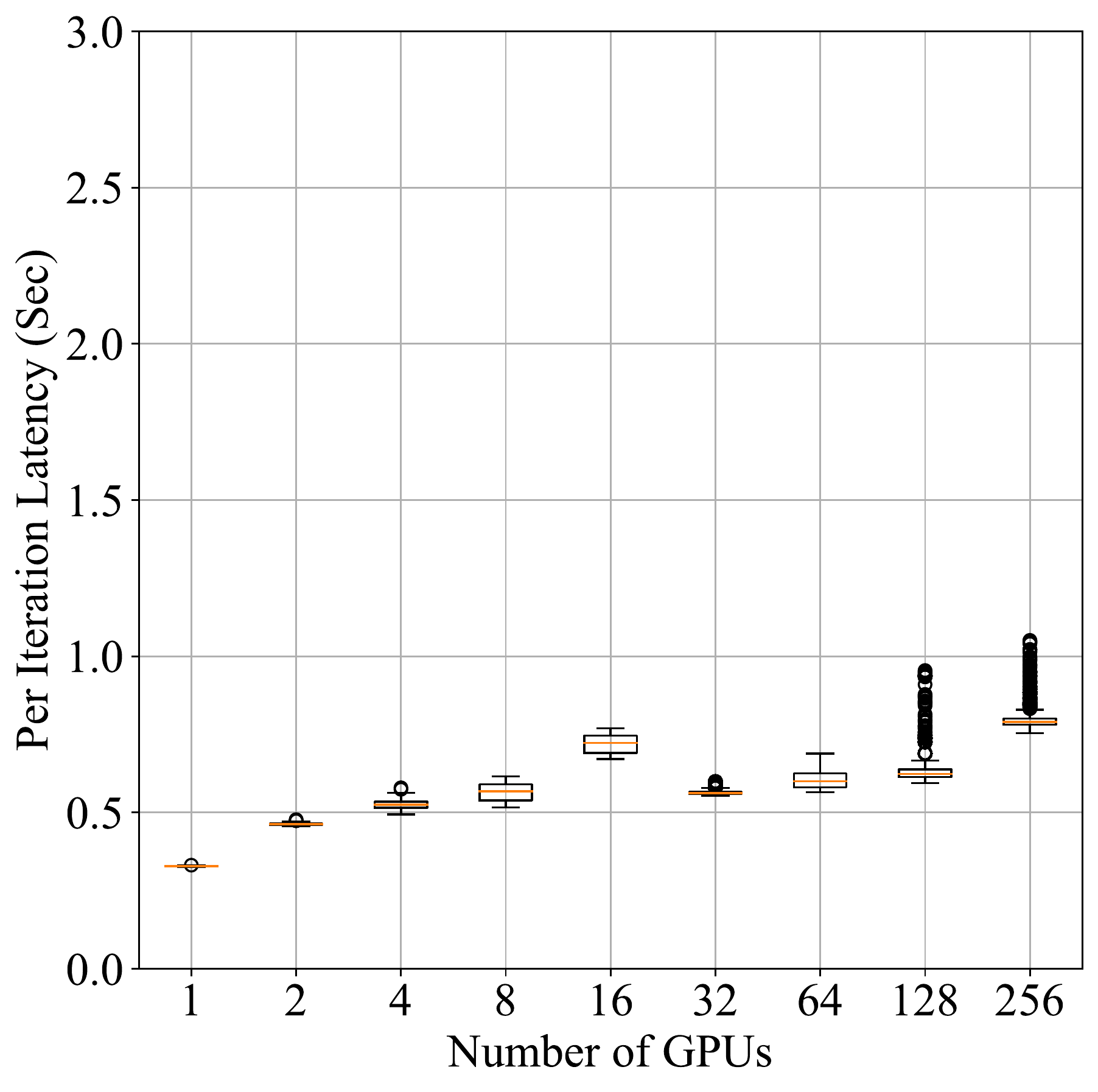}
  (c) BERT on NCCL
\end{minipage}
\begin{minipage}[c]{0.245\textwidth}
  \centering
  \includegraphics[width=\linewidth]{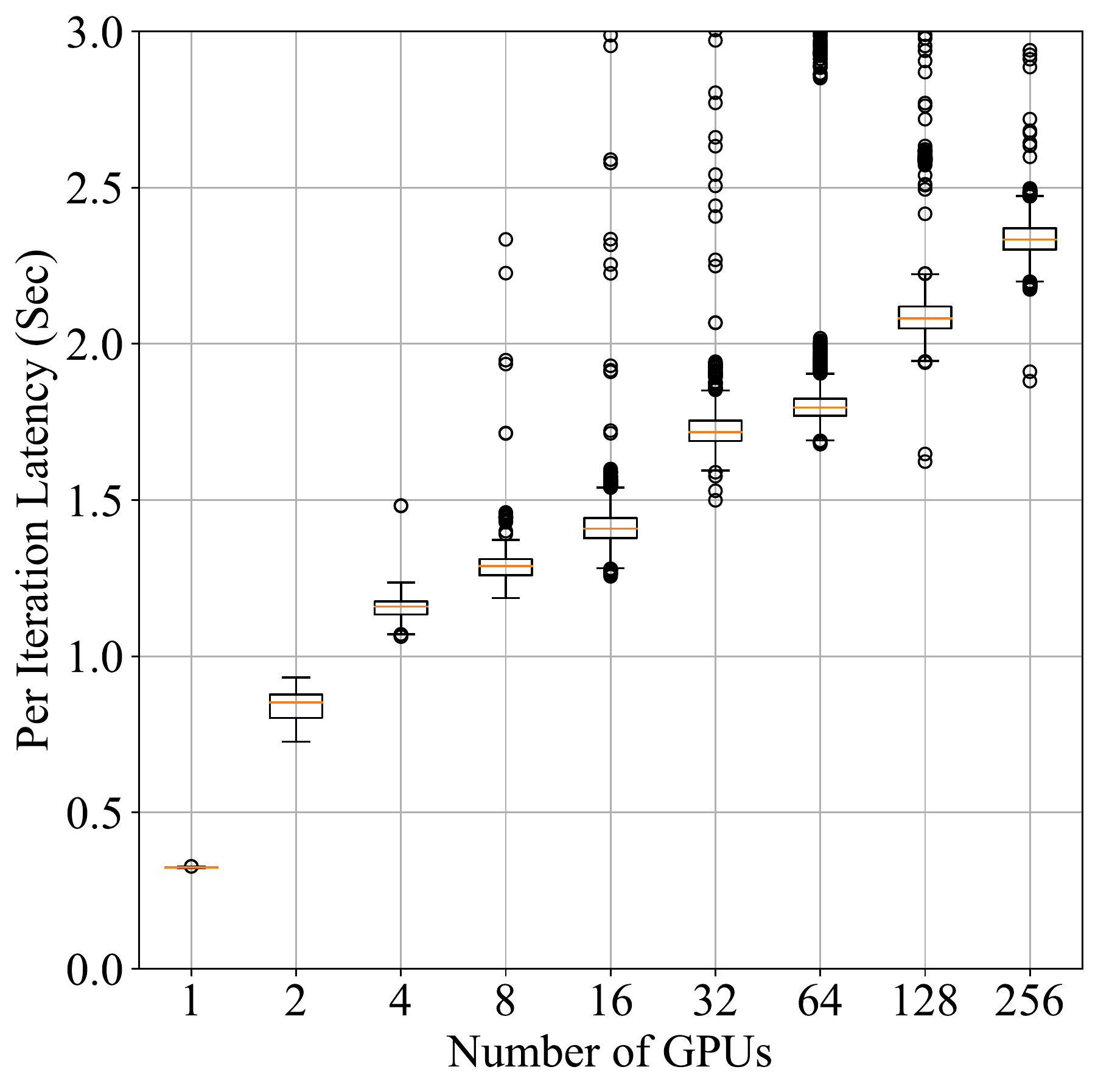}
  (d) BERT on Gloo
\end{minipage}
\vspace{-1em}
\caption{Scalability}\label{fig:scalability}
\vspace{-0.8em}
\end{figure*}

To understand the scalability of \code{DDP}, we measure per iteration training latency of ResNet50 and BERT using NCCL and Gloo backend on up to 256 GPUs in the shared entitlement. Results are presented in Fig.~\ref{fig:scalability}. The X-axis is the number of GPUs, and Y-axis the latency. Figure~\ref{fig:scalability}~(a) shows that the per iteration latency steadily increases as it scales out. Using 256 GPUs leads to 100\% slow down in each iteration compared to local training, meaning that the real scaling factor is 256 $\times$ 50\% = 128. With the BERT model, the per-iteration latency significantly increases due to the larger model size. Another observation is that the 16-GPU case suffers a longer per-iteration delay compared to the 32-GPU case in Figure~\ref{fig:scalability}~(c). We suspect this is because either the 16-GPU experiments were on a slow or congested link or there are other workflows in the shared entitlement competing for resources with our job. Fig.~\ref{fig:scalability}~(b) and (d) show the results for Gloo backend and the per-iteration slowdown is about 3X for ResNet and 6X for BERT when using 256 GPUs. The deteriorated training speed with larger model sizes indicates that the network is the bottleneck resource when using Gloo backend in this experiment.

\begin{figure*}[!htb]
\begin{minipage}[c]{.5\textwidth}
\begin{minipage}[c]{0.5\textwidth}
  \centering
  \includegraphics[width=\linewidth]{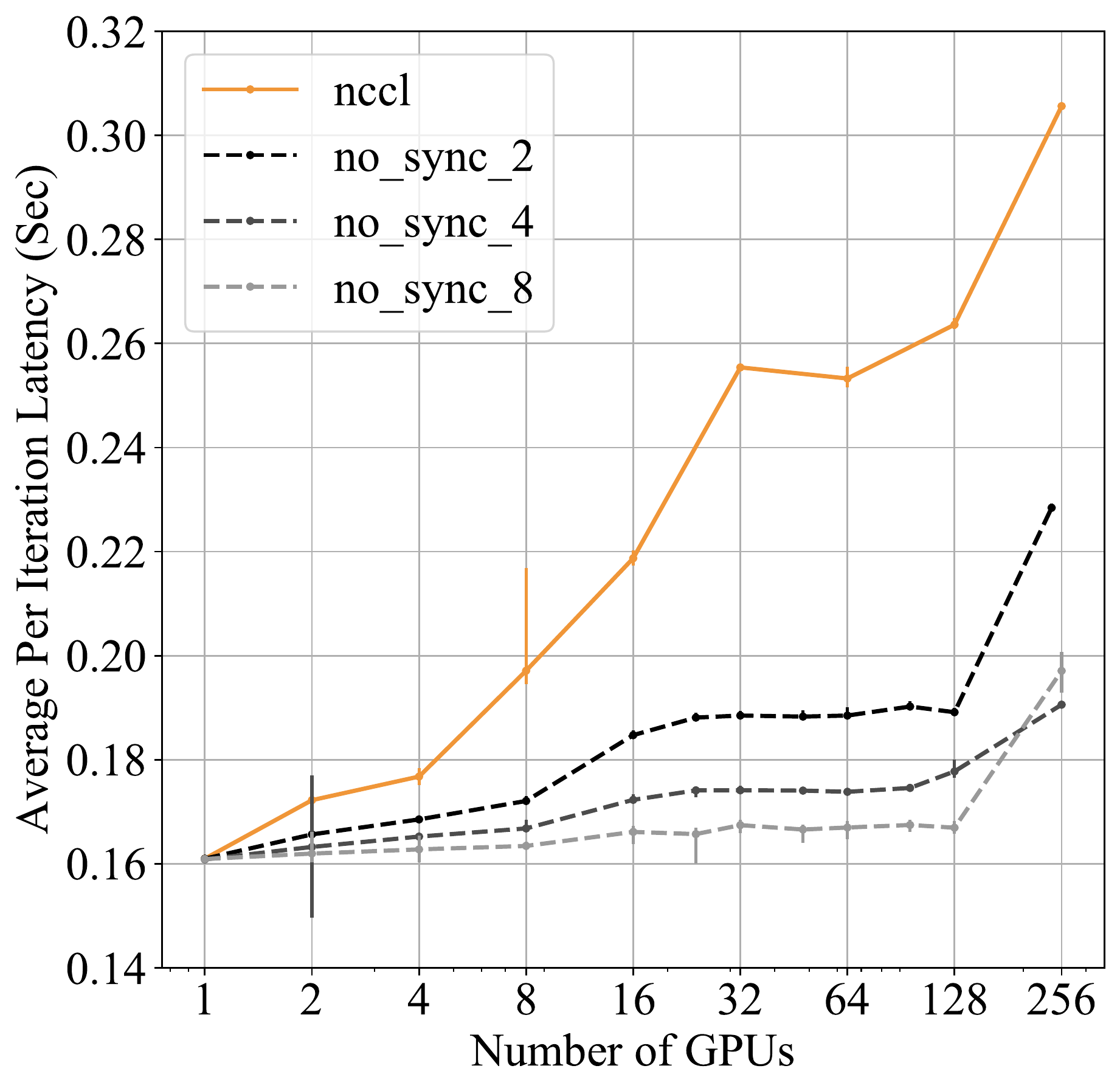}
  (a) ResNet50 on NCCL
\end{minipage}
\begin{minipage}[c]{0.5\textwidth}
  \centering
  \includegraphics[width=\linewidth]{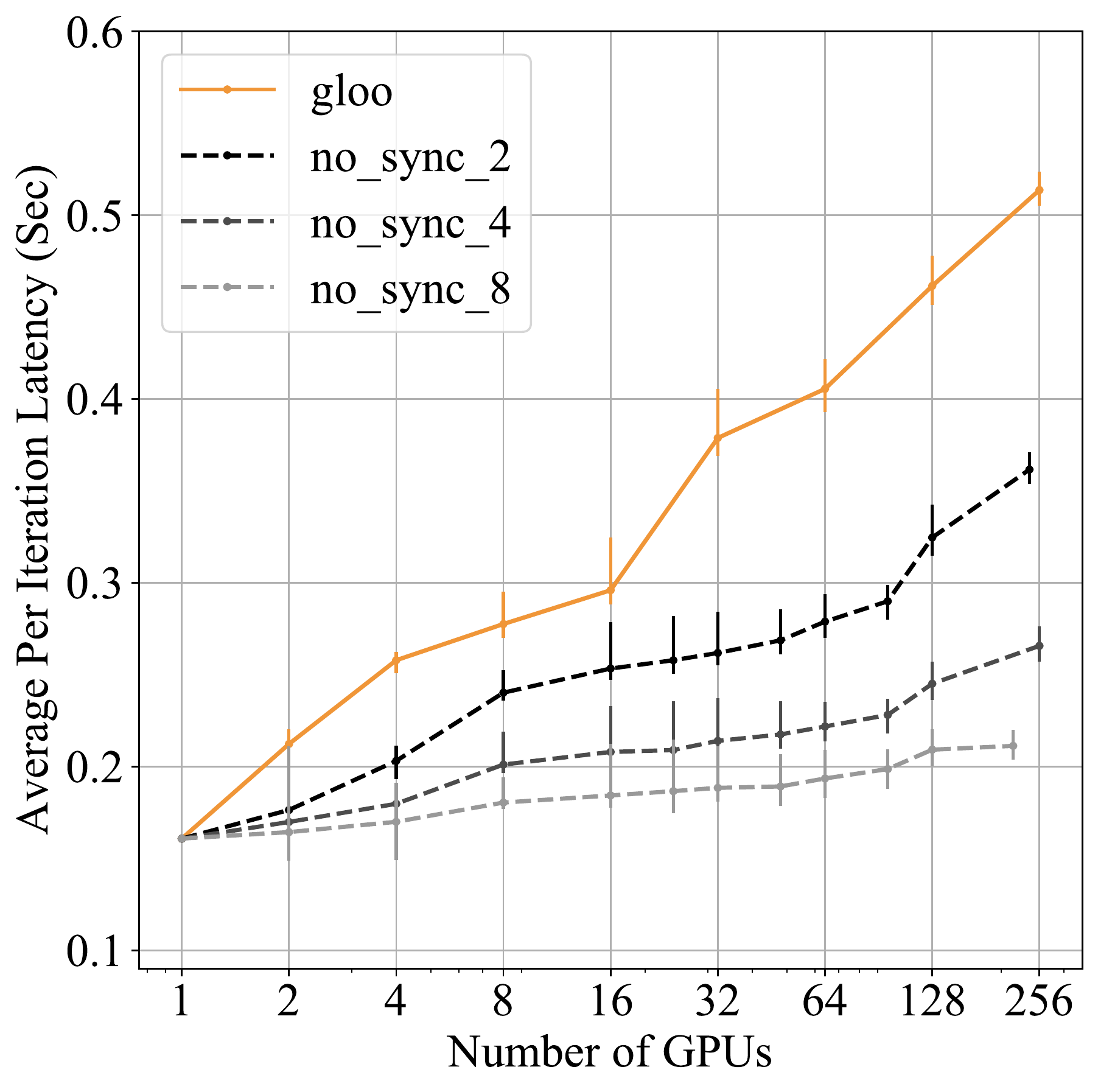}
  (b) ResNet50 on Gloo
\end{minipage}
\vspace{-1em}
\caption{Skip Gradient Synchronization}\label{fig:no_sync}
\end{minipage}
\begin{minipage}[c]{.5\textwidth}
\begin{minipage}[c]{0.5\textwidth}
  \centering
  \includegraphics[width=\linewidth]{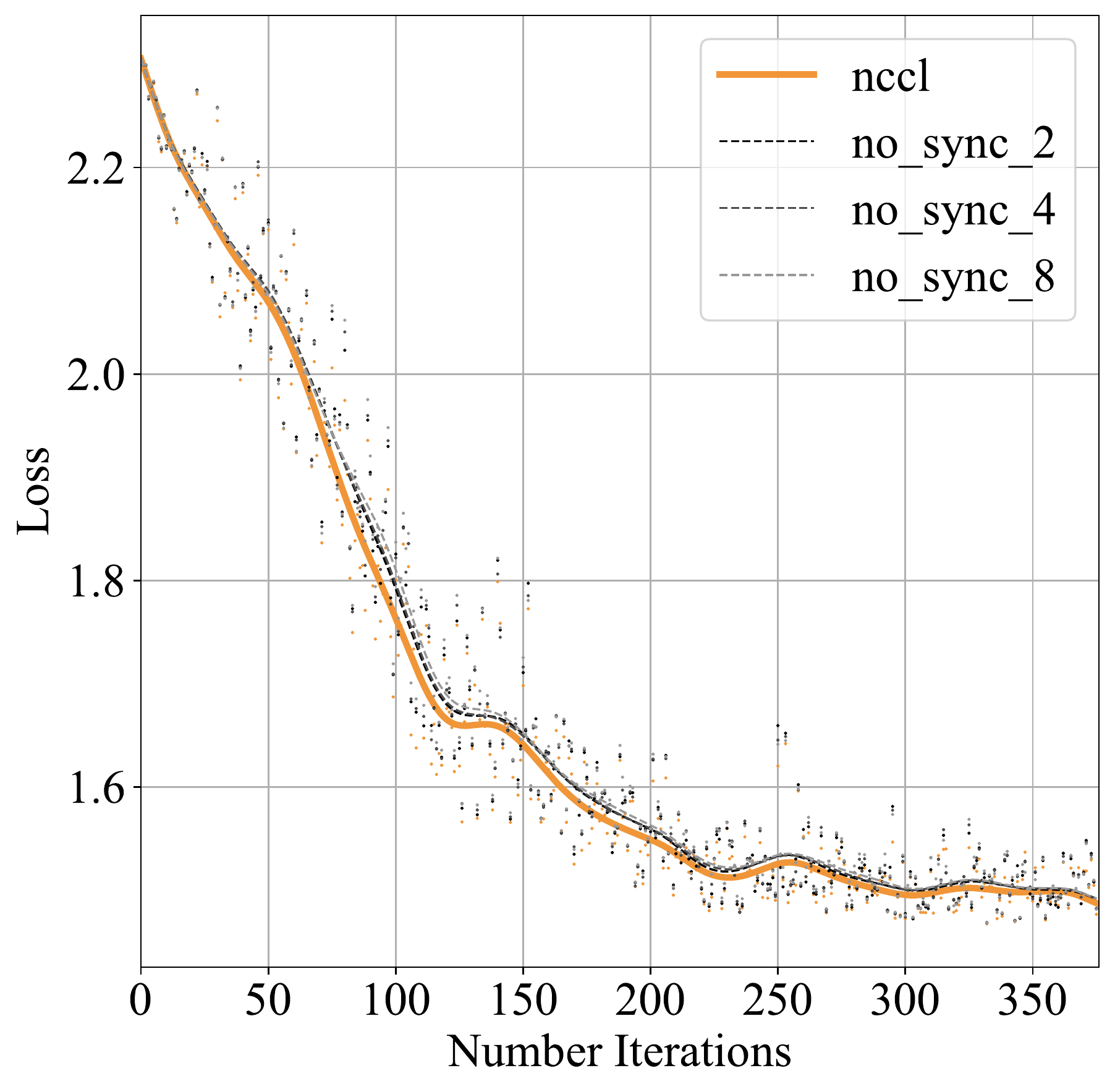}
  (a) Batch Size = 8
\end{minipage}
\begin{minipage}[c]{0.5\textwidth}
  \centering
  \includegraphics[width=\linewidth]{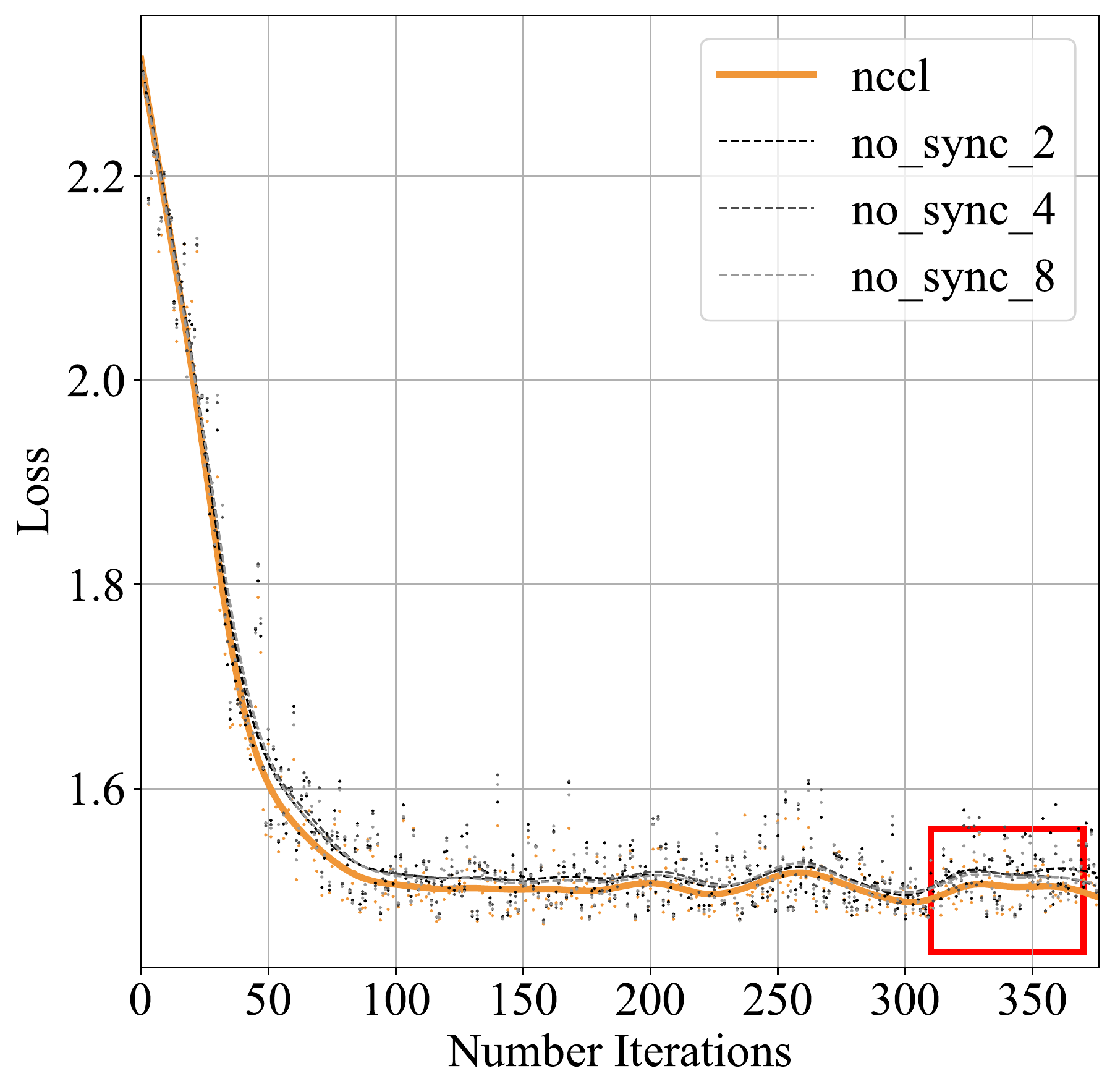}
  (b) Batch Size = 256
\end{minipage}
\vspace{-1em}
\caption{Accuracy with Skipping Synchronization}\label{fig:accuracy}
\end{minipage}
\vspace{-0.8em}
\end{figure*}

\begin{figure*}[!htb]
\begin{minipage}[c]{0.245\textwidth}
  \centering
  \includegraphics[width=\linewidth]{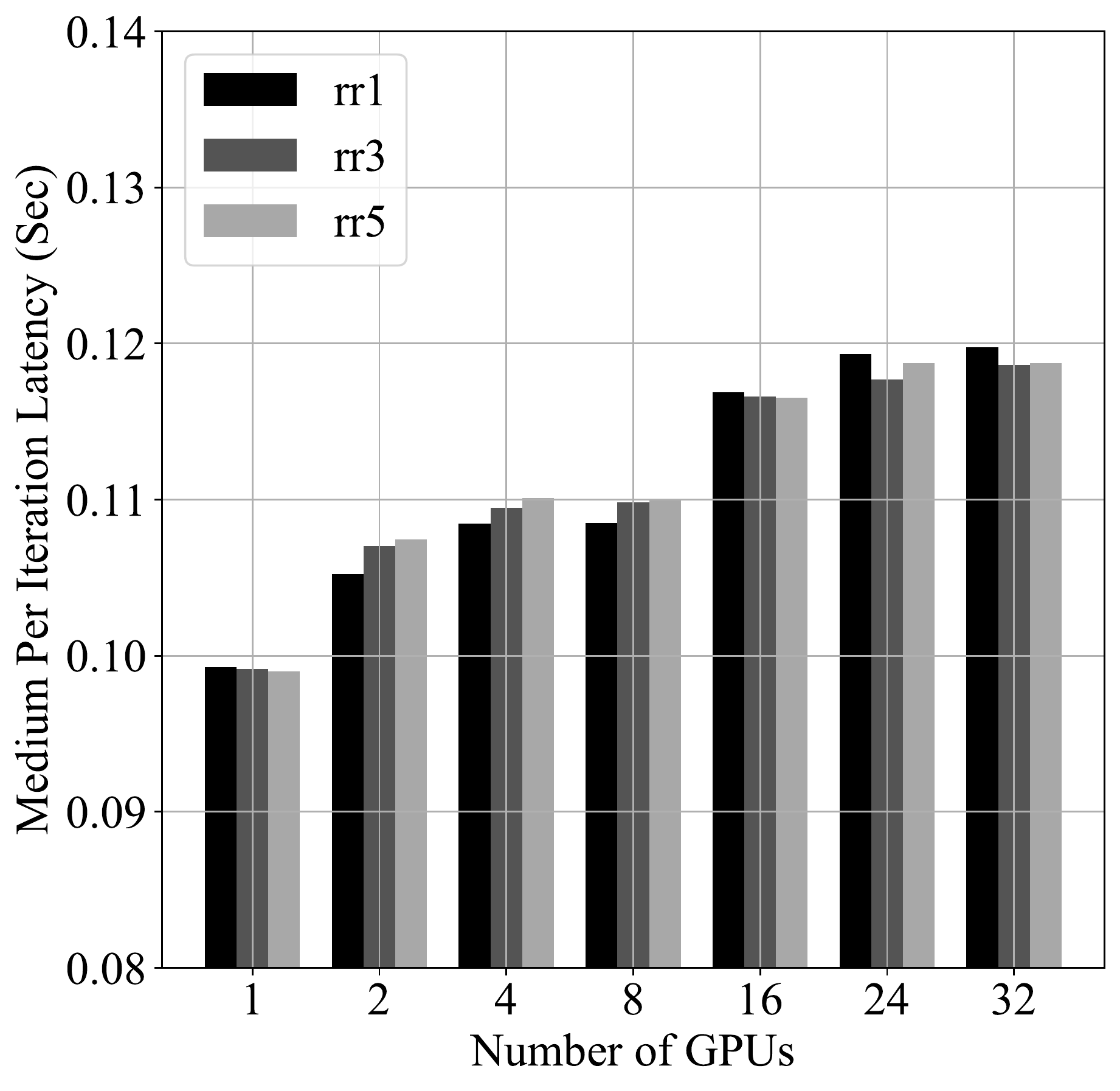}
  (a) ResNet50 on NCCL
\end{minipage}
\begin{minipage}[c]{0.245\textwidth}
  \centering
  \includegraphics[width=\linewidth]{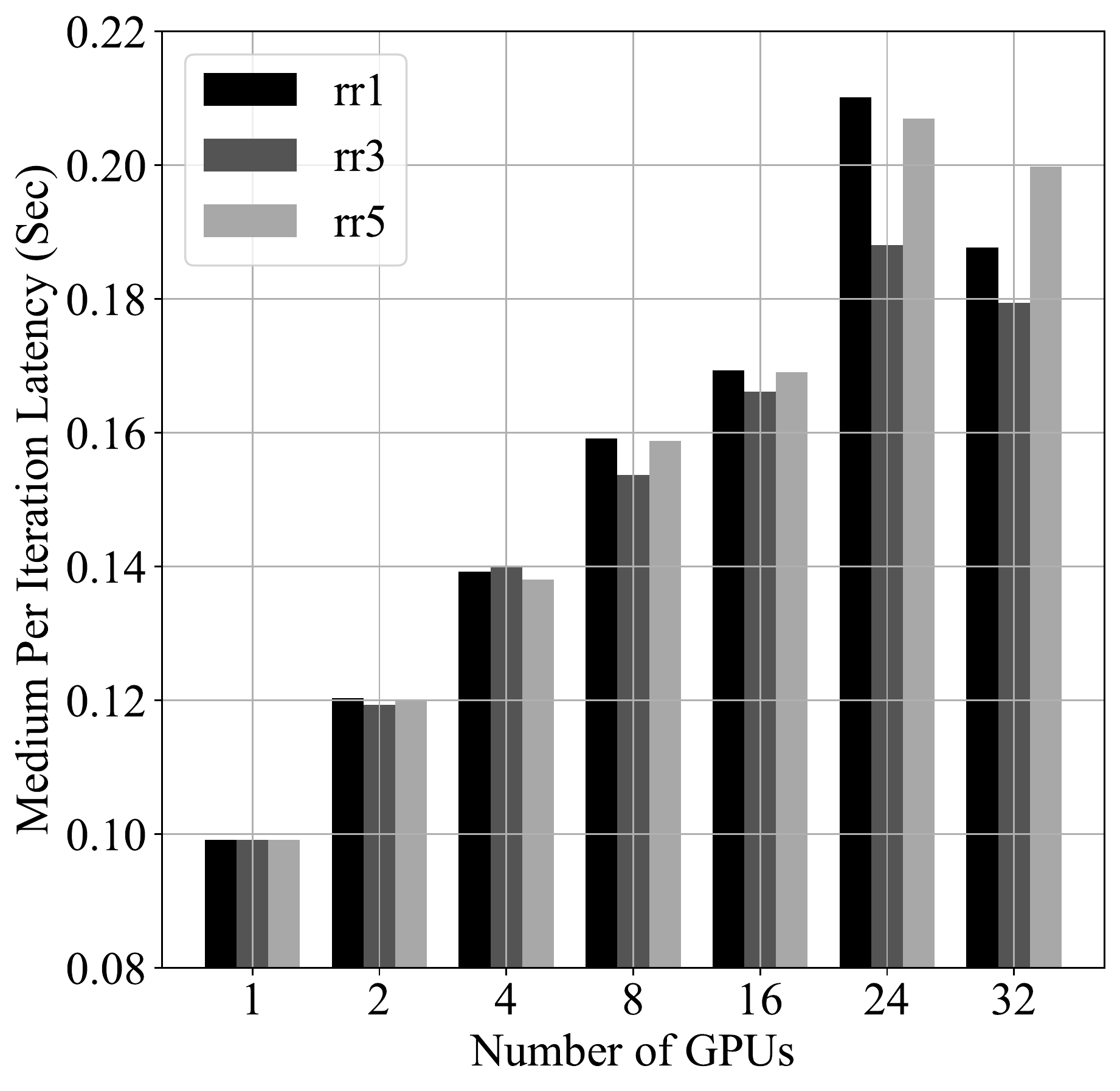}
  (b) ResNet50 on Gloo
\end{minipage}
\begin{minipage}[c]{0.245\textwidth}
  \centering
  \includegraphics[width=\linewidth]{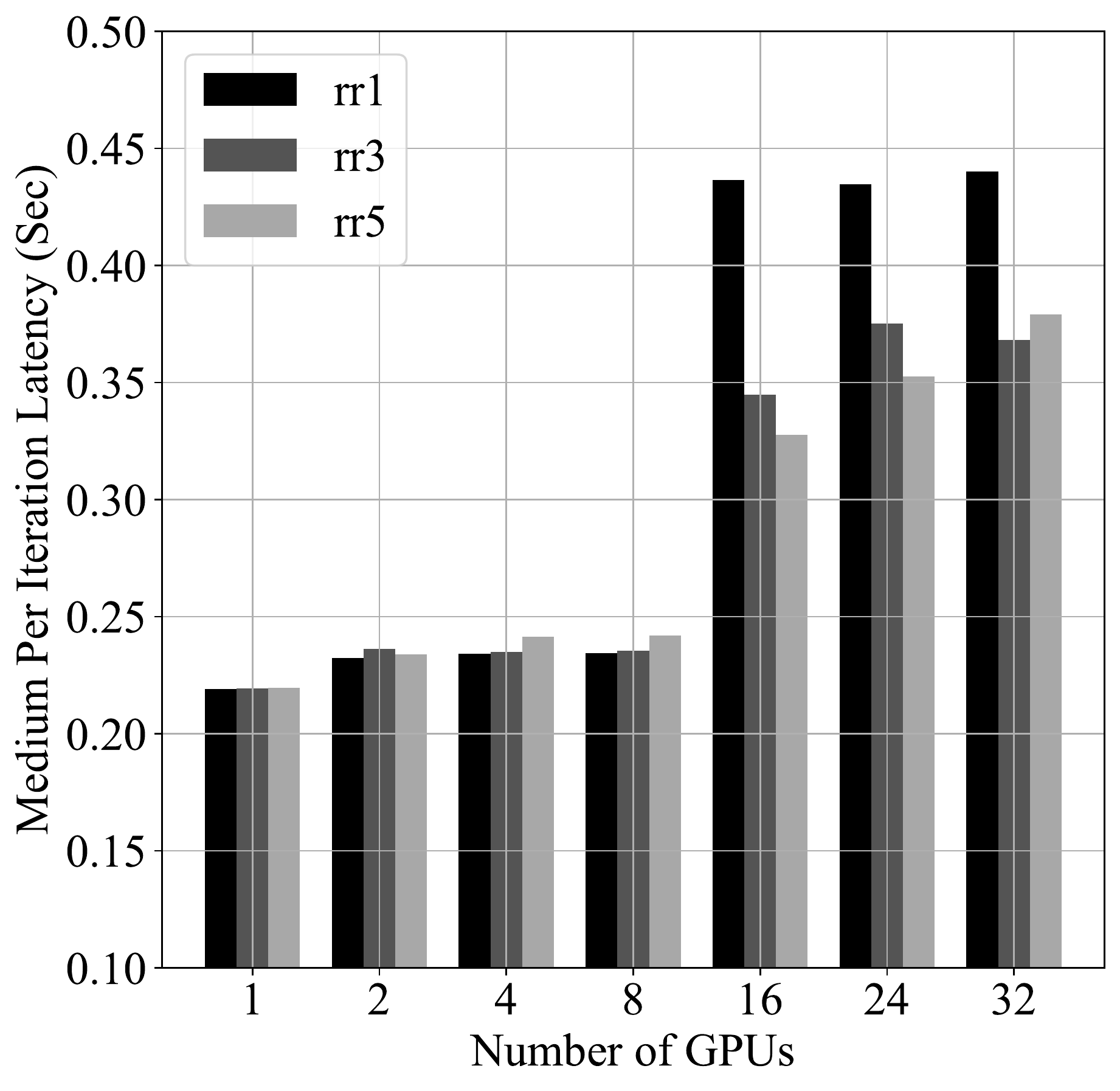}
  (c) BERT on NCCL
\end{minipage}
\begin{minipage}[c]{0.245\textwidth}
  \centering
  \includegraphics[width=\linewidth]{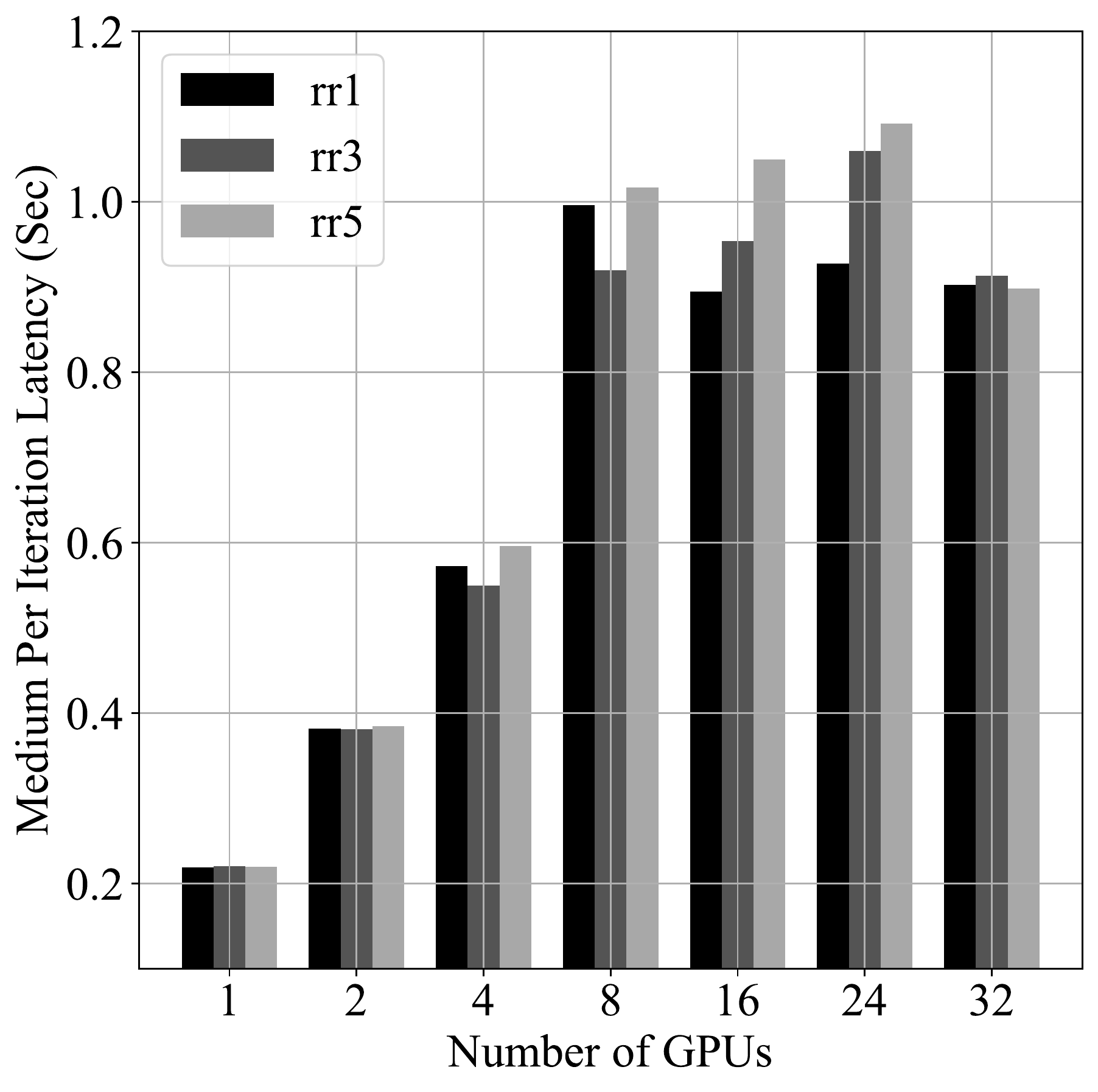}
  (d) BERT on Gloo
\end{minipage}
\vspace{-1em}
\caption{Round-Robin Process Group}\label{fig:rr}
\vspace{-1em}
\end{figure*}

In general, scaling out to more GPUs slows down individual iterations. One option to mitigate the overhead is skipping gradient synchronizations, \emph{i.e.}, perform gradient reduction every $n$ iterations. This approach helps to considerably reduce the amortized latency. Fig.~\ref{fig:no_sync} depicts the average per iteration latency for conducting gradient reduction every 1, 2, 4, and 8 iterations. To visually compare the effectiveness of this method, we consolidated different skipping configurations for the same model and backend combination into the same figure. 
ResNet50 on NCCL and Gloo sees 38\% and 57\% speed up with 256 GPUs when conducting gradient sync every 8 iterations. There is a sudden jump in delay with NCCL backend when scaling from 128 to 256 and this occurs to all experiments shown in this figure. We believe this is caused by slow or congested links among some of those 256 nodes which are not included in the 128-GPU experiments.
Besides the per iteration latency, it's also crucial to measure the convergence speed to verify if the acceleration might be erased by convergence slowdown. The experiments use MNIST~\cite{mnist} dataset to train the ResNet. The learning rate is set to 0.02 and the batch size is 8. Results are plotted in Fig.~\ref{fig:accuracy}~(a), which only contains the measurements for NCCL backend as the communication layer does not change the convergence speed. X-axis is the number of iterations and Y-axis the loss. Please note that the goal of this experiment is not developing the best model for MNIST, instead, it only aims to show the impact of skipping synchronization on the model convergence. The raw loss data oscillate severely, which are presented by the tiny dots. Directly connecting them into a line would result in the last curve covering all previous drawn ones, making them less visible. Therefore, we apply an order 3 low pass filter by using \code{filtfilt} from SciPy~\cite{scipy} and plot the smoothed loss curve. The figure confirms that using \code{no\_sync} in this case only leads to negligible exacerbation to the convergence speed. However, we must emphasize that the impact of \code{no\_sync} could depend on the configuration. Fig.~\ref{fig:accuracy}~(b) shows similar measurements by replacing batch size to 256 and learning rate to 0.06. As highlighted by the red box in the right bottom corner, \code{no\_sync} hurts the final training loss. It is because large batch size and \code{no\_sync} cause more gradients to be accumulated between consecutive communications and optimizer steps, which implicitly requires using a smaller learning rate. In summary, when skipping synchronizations properly, \code{DDP} attains near linear scalability with negligible accuracy penalty. 

\subsection{Round-Robin Process Group}


Another technique to speed up training is to use multiple process groups to work around subtle intrinsic concurrency limitations in process group backend implementations. The concurrency limitations could come from NCCL streams or Gloo threads, depending on the type of the backend, which might prevent one process group instance to fully utilize all link bandwidth. The PyTorch distributed package supports composing a Round-Robin process group with multiple NCCL or Gloo process groups, which dispatches collective communications to different process group instances in Robin-Robin order. Fig.~\ref{fig:rr} plots the per iteration latency of Round-Robin process group using 1, 3, and 5 NCCL or Gloo process groups, where \code{rrx} stands for \textbf{R}ound-\textbf{R}obin with \textbf{x} process group instances. ResNet50 on NCCL backend sees negligible differences with different amounts of process groups, meaning that for relatively small models like ResNet50, bandwidth is not the bottleneck resource. Noticeable difference can be observed in ResNet50 on Gloo, where \code{rr3} consistently outperforms \code{rr1}. The most prominent acceleration occurs in BERT model with NCCL backend, where \code{rr3} achieves 33\% speedup compared to \code{rr1} on 16 GPUs, revealing that one NCCL group is incompetent to saturate the link capacity.

\section{Discussion}\label{sec:discussion}

This section discusses lessons learned from our experiments and past experiences. We then present several ideas for future improvements. 

\subsection{Lessons Learned}

Distributed data parallel training is a conceptually simple or practically subtle framework. There are various techniques to improve its speed, creating a complex configuration space. Based on our observations, there is no single configuration that would work for all use cases, as it would highly depend on the model size, model structure, network link bandwidth, etc. However, on individual configuration dimensions, we summarized intuitions to help application developers to quickly navigate to a small range which likely contains the optimal solution for a given use case. The specific value of the configuration would require empirical measurements for every different deployment. 

\begin{itemize}
    \item \textbf{Communication Backend}: NCCL is considerably faster than Gloo in most use cases. When available, applications should seek to use NCCL as the primary collective communication backend.
    \item \textbf{Bucket Size}: Both excessively small or large bucket sizes are detrimental to communication performance. The optimal value lives in between and depends on the type of communication backend employed. The optimal bucket sizes are likely to increase with the size of the model in a sub-linear manner. 
    \item \textbf{Resource Allocation}: There is a significant slowdown with NCCL backend when scaling models across machine boundaries, if the bandwidth across machines is considerably lower than that between same-machine GPUs. In such cases, it is recommended to keep the \code{DDP} group within the same machine. If the training requires larger scale, developers can explore enabling \code{no\_sync} mode if it attains acceptable convergence speed. 
\end{itemize}

\subsection{Future Improvements}

While we implement and maintain the \code{DDP} package, several ideas for improvements popped up. This section discusses the basic ideas behind those improvements. 



\subsubsection{Gradient Order Prediction}

Although \code{DDP} cannot deterministically detect the backward computation order on all parameters at construction time, the order usually does not change that often in practice. One viable solution is to trace the backward order using autograd hooks and update parameter to bucket mapping accordingly. As bucket re-allocation will introduce noticeable overhead, it should be conducted infrequently. Given the existing complexities in \code{DDP}, tracing overhead should be negligible. Nevertheless, if there are disparities among tracing results from different iterations, additional complexities will be necessary to reach a consensus. 

\subsubsection{Layer Dropping}

One technique to accelerate training and avoid overfitting is to randomly drop layers during the forward pass~\cite{fan2019reducing}. This works well with local training. As every forward pass would build a new autograd graph, those skipped layers will not participate in the backward pass either. This idea also works with \code{DDP}, because parameters in skipped layers can be marked as ready in the forward pass and \code{DDP} will not wait for their autograd hooks during the backward pass. Although \code{DDP} would produce the correct result, this technique alone is inadequate to accelerate distributed data parallel training the same way as local training due to the fixed parameter-to-bucket mapping. As \code{AllReduce} uses a bucket as the minimum granularity, it cannot judiciously react to vacancies in buckets (\emph{i.e.}, skipped layers or parameters). Consequently, regardless of how the forward pass skips layers, there is always the same amount of data to be communicated across the wire during the backward pass. Besides, \code{DDP} cannot afford the luxury to adjust all buckets to cooperate with randomly skipped layers, as that would result in unacceptable memory allocation overhead. To tackle this problem, one solution is to keep bucket buffers intact but modify parameter-to-bucket mappings accordingly. Another option is to perform layer skips at the bucket level, \emph{i.e.}, \code{DDP} can map layers instead of parameters to buckets and all processes skip the same bucket in the same iteration. Both options require extra coordination across all \code{DDP} processes, which can be implemented by using the same random seed or having an authority process to broadcast the plan.

\subsubsection{Gradient Compression}

Another potential improvement for \code{DDP} is to reduce the volume of data for communication by compressing gradients. The absolute value of gradients are usually small, which might not require \code{float32} or \code{float64} types. Current \code{DDP} implementation always uses the parameter type as the gradient type that can become an overkill especially when the model is approaching convergence. In this case, \code{DDP} would benefit from adaptive compression levels by only communicating gradients with the necessary precision. Some recent research work~\cite{seide2014} even proposes more aggressive compression schemes, where by trading a tiny amount of model accuracy, applications can significantly accelerate distributed training by communicating just 1 bit for each gradient. 


\section{Related Work}\label{sec:related}

Distributed training algorithms can be categorized into different types from different perspectives. Below are three popular categorizations.

\begin{itemize}
    \item Synchronous update vs Asynchronous update: With the former, all model replicas can use \code{AllReduce} to collectively communicate gradients or parameters, while the asynchronous scheme employs P2P communication to update gradients or parameters independently.
    \item Cross-iteration vs Intra-iteration:  Cross-iteration parallelism (e.g., pipeline parallelism) allows the lifetime of multiple iterations to overlap with each other, while intra-iteration scheme focuses on parallelizing training within one iteration. 
    \item Data parallel vs Model parallel: Data parallel training distributes input data to multiple model replicas, while model parallelism divides the model into smaller pieces, which is especially helpful when the model is too large to fit in one device or machine.
\end{itemize}

Table~\ref{tab:related} summarizes some recent distributed training solutions by marking which scheme they can support. Besides advances in training schemes, prior work has also explored different communication algorithms, including tree-based \code{AllReduce}~\cite{nccl:tree_reduce}, heterogeneity-aware interconnection structure~\cite{wang2019blink}, and \code{AllReduce} decomposition~\cite{cho2019blueconnect}. As this paper focuses on \code{DDP}, the remainder of this section only elaborates and compares closely related techniques, \emph{i.e.}, Synchronous, Intra-iteration, and Data parallel training schemes. 

The techniques presented in this paper were first implemented and released in PyTorch v1.1. Similar computation-communication overlap techniques are also introduced in TensorFlow v2.2 as the Multi Worker Mirrored Strategy~\cite{tf-ddp}. This technique is researched in academia as well. GradientFlow~\cite{sun2019gradientflow} combines bucketing \code{AllReduce} with skipping parameter synchronizations. Compared to PyTorch \code{DDP}, instead of skipping the entire synchronization step in one iteration, GradientFlow selectively communicates a subset of gradients. Although this strategy helps to reduce communication overhead for gradients, it requires an additional communication phase to attain consensus on which gradients to synchronize. As a result, the overhead incurred to acquire consensus might overshadow the speedup achieved in gradient synchronizations, especially for small models or large network round-trip delays.

Another approach to speeding up distributed training is preempting and prioritizing communications based on the order of downstream computations. Jayarajan \emph{et al.}~\cite{sosp:19:priority} proposed to prioritize gradient synchronizations and parameter updates based on the forward order instead of the backward order, meaning that gradient buckets containing the initial layers should receive higher priorities than those in the final layers. Communications should still start from final layer gradients, as they will become ready earlier, but higher priority gradients (\emph{i.e.}, in initial layers) can preempt lower priority ones. This design allows the forward pass in the next iteration to start sooner, even before finishing gradients communications in the previous iteration, creating more opportunities to overlap computations and communications. ByteScheduler~\cite{bytescheduler} explored scheduling communications for distributed data parallel training as well. However, instead of binding with a single framework, ByteScheduler works for multiple frameworks by inserting a common core scheduler between framework APIs and framework engines and uses per-engine plugins to intercept communication invocations. To integrate with PyTorch, ByteScheduler builds on top of Horovod~\cite{sergeev2018horovod} which launches communication in the optimizer. One downside of this approach is that, there is a hard barrier between the backward pass and the optimizer step. As a result, communication can only overlap with the next forward pass instead of the current backward pass. With dynamic graphs, the next iteration might touch a different set of parameters, which would invalidate the schedule derived from the previous iteration. PACE~\cite{baopreemptive} computes the optimal communication schedule and implements preemption by segmenting primitive \code{AllReduce} operations into smaller pieces. Although segmenting can indeed mimic preemption, it will on the other hand hurt the total communication time as we have seen in Fig.~\ref{fig:allreduce}. A more efficient approach would be to natively support prioritization in the communication libraries (e.g., NCCL and Gloo).

\begin{center}
\begin{table}
\begin{center}
\begin{tabular}{ |c|cccccc| }
\hline
Scheme & S & A & C & I & D & M\\
\hline
PT DDP~\cite{pt-ddp} & $\surd$ & & & $\surd$ & $\surd$ & \\
\hline
PT RPC~\cite{rpc} &  & $\surd$  & $\surd$ & $\surd$  & $\surd$  & $\surd$ \\
\hline
TF MultiWorkerMirrored~\cite{tf-ddp} & $\surd$ & & & $\surd$ & $\surd$ & \\
\hline
TF ParameterServer~\cite{tf-ps, ps} & & $\surd$  & & $\surd$ & $\surd$ & $\surd$ \\
\hline
Mesh TensorFlow~\cite{shazeer2018mesh} & $\surd$ &  & & $\surd$ & $\surd$ & $\surd$ \\
\hline
GPipe~\cite{gpipe} & $\surd$ &  & $\surd$ &  & & $\surd$ \\
\hline
Horovod~\cite{sergeev2018horovod} & $\surd$ & & & $\surd$ & $\surd$ & \\
\hline
GradientFlow~\cite{sun2019gradientflow} & $\surd$ & & & $\surd$ & $\surd$ & \\
\hline
SlowMo~\cite{wang2019slowmo} & & $\surd$  & & $\surd$ & $\surd$ & \\
\hline
PipeDream~\cite{narayanan2019pipedream} & $\surd$ &  & $\surd$ & $\surd$  & $\surd$  & $\surd$ \\
\hline
ZeRO~\cite{rajbhandari2019zero} & $\surd$ &  & & $\surd$  & $\surd$  & $\surd$ \\
\hline
Parallax~\cite{parallax} & $\surd$ & $\surd$ & & $\surd$  & $\surd$  & $\surd$ \\
\hline
ByteScheduler~\cite{bytescheduler} & $\surd$ & & $\surd$  & $\surd$  & $\surd$  & \\
\hline
TicTac~\cite{hashemi2018tictac} & & $\surd$ & & $\surd$  & $\surd$  & $\surd$  \\
\hline
PACE~\cite{baopreemptive}  & $\surd$ &  & & $\surd$  & $\surd$  &   \\
\hline
\end{tabular}
\end{center}
\caption{Distributed Training Solutions: \normalfont{ Six schemes are \textbf{\b{S}}ynchronous-Update vs \textbf{\b{A}}synchronous-Update, \textbf{\b{C}}ross-Iteration vs \textbf{\b{I}}ntra-Iteration, \textbf{\b{D}}ata-Parallel vs \textbf{\b{M}}odel-Parallel}\vspace{-1em}}\label{tab:related}
\end{table}
\end{center}

The mixture of different parallelism scheme fosters even more powerful training paradigms. Mesh-TensorFlow~\cite{shazeer2018mesh} combines data parallelism with model parallelism. It vertically divides some layers by dimensions and replicating other layers where the given dimension is absent. ZeRO~\cite{rajbhandari2019zero} also combines data parallelism with model parallelism, but with minimum model replication to support fast training on super large models. The authors observed that main memory consumption contributors are input data, model parameters, gradients, optimizer states, and activations. Splitting input data is trivial. However, model parameters and activations are compulsory ingredients for backward passes. ZeRO addressed this problem by partitioning parameters, gradients, and optimizer states on each \code{DDP} instance. Parameters are broadcast from the owner \code{DDP} instance to all others when necessary. Activations are recomputed during the backward pass. Compared to PyTorch \code{DDP}, ZeRO can scale to much larger models as each process only needs to maintain a small partition of the model. The high scalability is achieved by sacrificing the training speed, as the additional re-computation, broadcast, and gather would introduce considerable overhead. Hence, applications can choose which techniques to use based on the size of the given model and available resources. PipeDream~\cite{narayanan2019pipedream} employs a different approach where the model stack is decomposed into multiple stages, where data parallelism is applied within one stage and pipeline with model parallelisms govern the workload across stages. One subtle detail is that to attain high training speed, PipeDream slightly sacrifices accuracy by using the latest gradients from multiple concurrent passes. Although the gradient might not be derived from the current parameter states, the authors show that this mismatch is tolerable in practice.  Parallax~\cite{parallax} explored a hybrid structure that combines parameter-server~\cite{ps} and collective communications. Models are partitioned based on  sparsity, where dense parameters are communicated using \code{AllReduce} and sparse tensors are placed to parameter servers. This design avoids densifying sparse tensors and communicating empty values, which is especially helpful for NLP models.

\section{Conclusion}\label{sec:conclusion}

This paper explained the design and implementation of the distributed data parallel module in PyTorch v1.5, and conducted performance evaluations on NCCL and Gloo backend using ResNet50 and BERT models. \code{DDP} accelerates training by aggregating gradients into buckets for communication, overlapping communication with computation, and skipping synchronizations. We also highlighted real-world caveats in gradient synchronization which are important for broad adoption.  Results showed that \code{DDP} with NCCL backend can achieve near-linear scalability on 256 GPUs when configured properly. The measurements also revealed that the backward pass in \code{DDP} is the most expensive step in training and requires efforts from both framework developers to enable optimization algorithms and application developers to empirically configure the knobs. Based on our observations, we shared lessons learned from serving a variety of application, discussed potential future improvements for distributed data parallel training, and enthusiastically encourage open source community to experiment with more novel ideas.

\newpage

\bibliographystyle{abbrv}
\bibliography{main}
\par\leavevmode

\end{document}